\documentclass{aa}
\usepackage{graphicx}
\usepackage{lscape}
\usepackage[varg]{txfonts}
\usepackage{natbib}
\usepackage{hyperref}
\usepackage{orcidlink}
\usepackage{CJK}

\hypersetup{
    colorlinks=true,
    linkcolor=blue,
    filecolor=blue,      
    urlcolor=blue,
    citecolor=blue
}

\begin{document}
\begin{CJK*}{UTF8}{gbsn}

\title{Searching for Galactic Red Supergiants with Gaia RVS Spectra}
   
\author{Zehao Zhang (张泽浩)\inst{1,2} \orcidlink{0000-0002-3828-9183}
        \and
        Biwei Jiang (姜碧沩)\inst{1,2} \orcidlink{0000-0003-3168-2617}
        \and
        Yi Ren (任逸)\inst{3}  \orcidlink{0000-0003-1218-8699}
        \and
        He Zhao (赵赫) \inst{4} \orcidlink{0000-0003-2645-6869}
        \and
        Ming Yang (杨明) \inst{5} \orcidlink{0000-0001-8247-4936}
        }
\institute{Institute for Frontiers in Astronomy and Astrophysics, Beijing Normal University, Beijing 102206, People's Republic of China \\
          \email{bjiang@bnu.edu.cn}
          \and
          School of Physics and Astronomy, Beijing Normal University, Beijing 100875, People's Republic of China 
          \and
          College of Physics and Electronic Engineering, Qilu Normal University, Jinan 250200, People's Republic of China 
          \and
          Purple Mountain Observatory and Key Laboratory of Radio Astronomy, Chinese Academy of Sciences, 10 Yuanhua Road, Nanjing 210033, People's Republic of China 
          \and
          Key Laboratory of Space Astronomy and Technology, National Astronomical Observatories, Chinese Academy of Sciences, Beijing 100101, People's Republic of China 
         }


\date{Received xx / Accepted xx}

\abstract{Red supergiants (RSGs) are essential to understanding the evolution and the contribution to the interstellar medium of massive stars. However, the number of identified RSGs within the Milky Way is still limited mainly due to the difficulty of measuring stellar extinction and distance. The release of approximately one million RVS spectra in Gaia DR3 presents new opportunity for identifying Galactic RSGs, because the equivalent width of the calcium triplet lines (EW(CaT)) in the spectra is an excellent indicator of stellar surface gravity. This work uses the RVS spectra with signal-to-noise ratio (SNR) greater than 100 to search for the Galactic RSGs. The dwarf stars and red giants are removed and the RSG candidates are selected by the location in the EW(CaT) vs. $BP-RP$ diagram. The early-type RSG candidates (K0-M2) are then identified by $BP-RP > 1.584$ and EW(CaT) $>$ 1.1 nm. To identify late-type RSG candidates (after M2), the criteria of the average equivalent widths of TiO in the XP spectra (EW(TiO)) $>$ 10 nm, the color index $K-W3 < 0.5$ and the period-amplitude sequence from Gaia DR3 LPV catalog are further applied to reduce the contamination of late-type red giants and asymptotic giant branch stars. This method yields 30 early-type (K0-M2) and 6196 late-type (after M2) RSG candidates, which is a significant increase to the present Galactic RSG sample. The application of this approach to the spectra with SNR $>$ 50 results in 48 early-type and 11,491 late-type RSG candidates. This preliminary analysis paves the way for more extensive research with Gaia DR4 when larger spectral datasets are expected to significantly enhance our understanding of Galactic RSG populations.}

\maketitle

\nolinenumbers

\section{Introduction}

Red supergiants (RSGs) are among the most massive and luminous stars in the Universe, representing a crucial evolutionary phase for stars with initial mass between approximately 8 and 40 $M_{\odot}$ \citep{1979ApJ...232..409H, 2003AJ....126.2867M}. RSGs are characterized by their immense size, cool temperature, complex light variability \citep{2006MNRAS.372.1721K, 2012ApJ...754...35Y, 2019ApJS..241...35R, 2020ApJ...898...24R, 2024ApJ...969...81Z}, and significant mass-loss rate \citep{2020AJ....160..145H, 2020MNRAS.492.5994B, 2021ApJ...912..112W, 2023A&A...676A..84Y, 2024AJ....167...51W, 2024A&A...681A..17D}, which have profound affects for their subsequent evolution and eventual fate as supernovae.

Identifying RSGs in nearby galaxies has recently made great progress with the fast development of observation and new method. In the Small and Large Magellanic Clouds (SMC and LMC), the parallax, proper motions and radial velocity (RV) measured by Gaia are highly reliable parameters to select the member stars \citep{2019A&A...629A..91Y, 2021ApJ...923..232R}. Though the Gaia astrometric parameters are accurate to select member stars, they are unavailable to most stars in the galaxies more distant than the SMC and LMC because of the limited sensitivity of Gaia. Instead, the color-color diagram (CCD) method is invented to select member stars. The very early attempt is carried out by \citet{1998ApJ...501..153M}. They use the $B-V/V-R$ diagram to separate the foreground dwarf stars from the member giant stars in that the $B$ band covers some metallic lines sensitive to surface gravity. \citet{2021ApJ...907...18R} improve this method by shifting the waveband to the near-infrared (NIR) $J-H/H-K$ diagram. The NIR CCD has the advantages of being consistent with the peak of the RSG spectral energy distribution (SED) and much less affected by interstellar extinction. After the foreground stars are removed, RSGs in an external galaxy can be easily identified by their high luminosity and red color in the color-magnitude diagram (CMD) since all the member stars are at almost the same distance. Consequently, 5498 and 3055 RSGs are identified in the M31 and M33 \citep{2021ApJ...907...18R}, 4823 and 2138 RSGs are identified in the LMC and SMC, and a total of 2190 RSGs are found in other 10 dwarf galaxies \citep{2021ApJ...923..232R}.

The identification of RSGs in the Galaxy lags behind. The Galactic RSGs distribute at various direction and distance so that no identical proper motion, parallax or RV can be used. Thus, the astrophysical parameters like luminosity and color index should be determined to find the RSGs. However, the inhomogeneous and heavy extinction leads to the measurement of both luminosity and intrinsic color index difficult (e.g. \citet{2010A&A...511A..18S}) because RSGs are located in the Galactic plane. Besides, the distance should be measured individually, while the perturbation of the photocenter of RSGs brings about additional uncertainty in the Gaia parallax \citep{2011A&A...528A.120C}. Instead of photometry, the identification of RSGs in the Galaxy relies heavily on spectroscopy. Indeed, the very early identification of RSGs in nearby galaxies also uses the spectral features and the RV derived from spectrum (see e.g. \citet{1988AJ.....96.1884H}).

Spectrum contains abundant information for stellar classification. In particular, the near-infrared spectra are suitable for late-type stars \citep{1991ApJS...77..417K, 1994A&AS..108..359G, 1997A&AS..123....5C}. Based on the spectral classifications by \citet{1978ApJS...38..309H} and \citet{1992A&AS...94..211G}, \citet{2005ApJ...628..973L} built a catalog of 74 Galactic RSGs, with their spectral types and effective temperature scales. Additionally, attention has been directed towards star clusters within the Milky Way, as RSGs are often found within OB associations. There are three massive RSGs-rich clusters named RSGC1, RSGC2, and RSGC3 located in a small region of the Galactic plane between $l = 24^\circ$ and $l = 29^\circ$ \citep{2006ApJ...643.1166F, 2007ApJ...671..781D, 2008ApJ...676.1016D, 2009A&A...498..109C, 2009AJ....137.4824A}, as well as several smaller clusters close to them \citep{2010A&A...513A..74N, 2011A&A...528A..59N, 2012A&A...547A..15N}. In some recent studies, \citet{2016A&A...595A.105D} used principal component analysis (PCA) and support vector machine (SVM) techniques to analyze the spectral features in the calcium triplet (CaT) region, establishing the criteria to distinguish supergiants from non-supergiants. Subsequently, they applied this method to identify 197 cool supergiants in the Perseus arm \citep{2018MNRAS.475.2003D}. \citet{2019AJ....158...20M} selected Galactic K-M type Class I stars and identified 889 RSG candidates, by considering the Gaia DR2 astrometric data, but only a small fraction of them are confirmed as true RSGs. Furthermore, \citet{2023A&A...671A.148M} identified 20 new RSGs through an analysis of Gaia DR3 GSP-Phot and GSP-Spec parameters in combination with BP/RP spectra. More recently, \citet{2024MNRAS.529.3630H} compiled a catalog of 578 highly probable and 62 likely RSGs. Despite the Mliky Way being estimated to host at least 5000 RSGs \citep{1989IAUS..135..445G}, the currently known sample is far smaller. This discrepancy is mainly due to the limitations of spectroscopic surveys, which typically detect only the bright sources, leading to the miss of faint sources. Spectroscopy is also generally less efficient than photometry. Fortunately, Gaia DR3 includes a vast number of medium-resolution RVS spectra \citep{2018A&A...616A...6S,2023A&A...674A...6S,2023A&A...674A..29R}. Besides, the space observation avoids the absorption lines of the atmosphere, which improves the accuracy of line measurements. It offers an unprecedented opportunity to revisit the population of Galactic RSGs from a new perspective.

This work aims to create a large catalog of Galactic RSGs through a comprehensive analysis of the RVS spectra. The paper is organized as follows: Section \ref{sec:data} introduces the RVS speactra and the process of re-normalization, as well as the selection of the initial sample. Section \ref{sec:RSG} describes the CaT characteristics of preliminary RSG samples, followed by the selection processes of early-type and late-type RSG candidates. We discuss the known Galactic RSGs, completeness and pureness in our sample, and present an outlook for Gaia DR4 in Section \ref{sec:discussion}. A summary is presented in Section \ref{sec:summary}.

\section{Data} \label{sec:data}
\subsection{RVS spectra} \label{sec:RVS_spectra}

Launched by the European Space Agency in 2013, the Gaia mission aims to create the most detailed three-dimensional map of our galaxy. Gaia's suite of instruments includes a highly precise astrometric detector, photometric facility, and the Radial Velocity Spectrometer (RVS, \citet{2016A&A...595A...1G}). The RVS is designed to obtain medium-resolution spectra in the wavelength range of 845-872 nm \citep{2018A&A...616A...6S}, also known as CaT region, which is suitable to determine the RV over a wide range of metallicity. This spectral region is very valuable for studying cool stars like RSGs, as it contains strong absorption lines that are sensitive to temperature, surface gravity, and metallicity \citep{2021A&A...654A.130C,2023A&A...674A..26C, 2023A&A...674A..28F}, in particular, CaT itself is an excellent indicator of surface gravity \citep{1989MNRAS.239..325D, 1994A&AS..103..279M, 1997A&AS..124..359M, 2001MNRAS.326..959C, 2001MNRAS.326..981C}.

Gaia DR3 includes 999,645 RVS spectra with a resolution of $R\sim 11150$. These spectra are wavelength-calibrated under vacuum conditions, and the majority has a signal-to-noise ratio (SNR) between 20 and 40 \citep{2023A&A...674A...1G}. They are primarily used for measuring stellar RV \citep{2023A&A...674A...5K} or for deriving GSP-Spec parameters with the Apsis pipeline \citep{2023A&A...674A..26C, 2023A&A...674A..28F, 2023A&A...674A..29R}. The RVS spectra cover the features for a wide range of stellar types, with early-type stars being dominated by hydrogen Paschen lines, while late-type stars exhibit more metallic and molecular lines (see Figure 6 of \citet{2023A&A...674A..28F}). The publicly released RVS spectra are normalized either by using their pseudo-continuum or by scaling with a constant (the latter for cool stars or spectra with low SNR, \citet{2023A&A...674A...1G}). For the majority of stars, the normalized flux is set to 1. However, due to the presence of TiO band in the spectra of late-type stars which erode the pseudo-continuum, the normalization process is often incorrect for M-type stars. As a result, their spectra exhibit a pronounced slope across the entire wavelength range (e.g., Figure 17 in \citet{2018A&A...616A...5C}). Given that Galactic RSGs are mostly of M-type (see Figure 5 of \citet{2012AJ....144....2L}), it is necessary to re-normalize the spectra.

\begin{figure*}[htp]
	\centering
     \includegraphics[width=\textwidth]{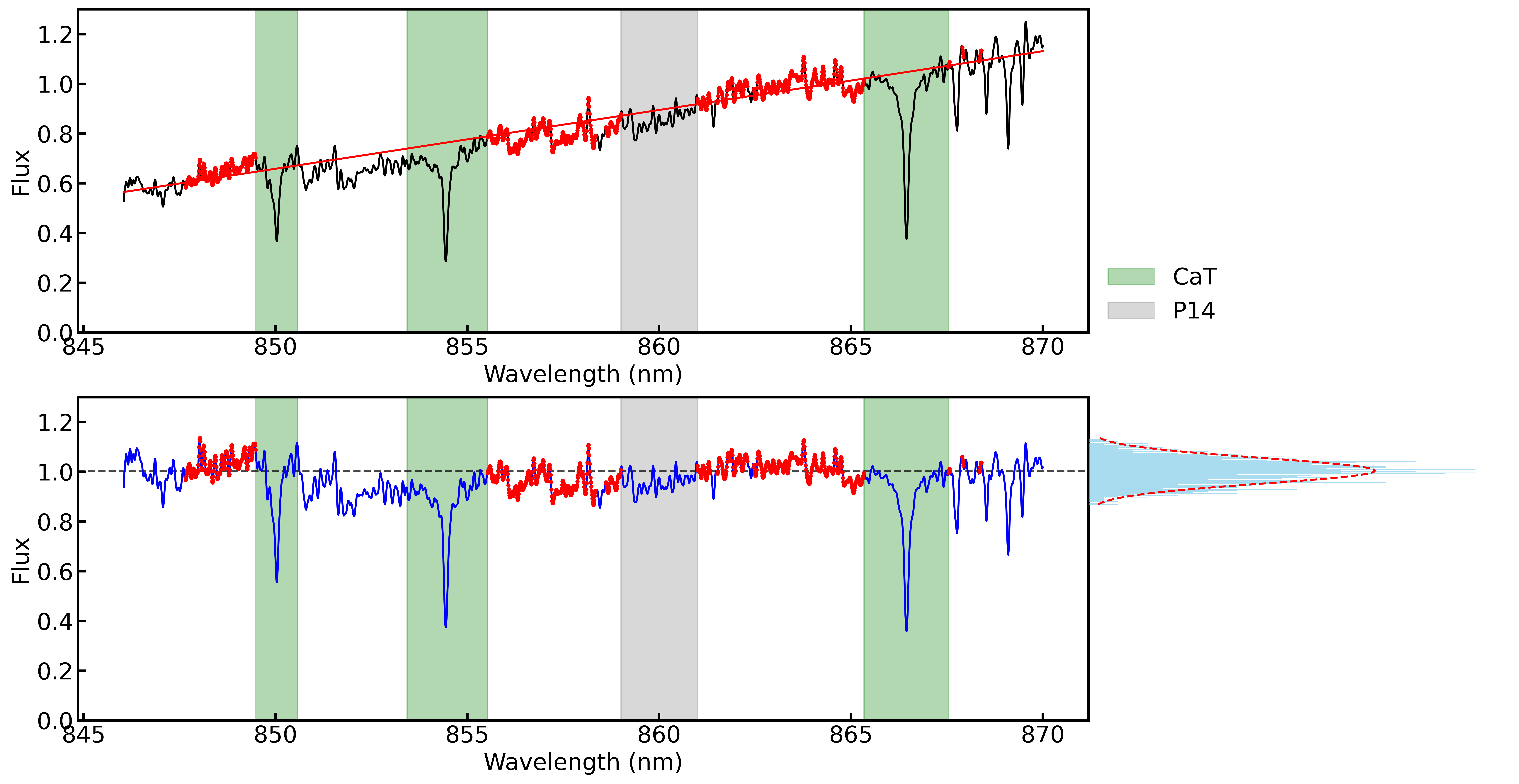}
	\caption{Example of RVS spectrum re-normalization ({\tt source\_id}: 5836392511305856128). The upper panel shows the original spectrum, and the solid red line is obtained by linear fitting of the red dots. The blue on the bottom panel is the renormalized spectrum, the histogram on the lower right is a distribution of red dots, and the black dashed line marks the mean of its Gaussian fitting, representing the continuum after re-normalization. The shaded colors are used to represent different absorption lines. \label{fig:spectrum}}
\end{figure*}

The steps for continuum re-normalization are as follows. First, the spectral regions with no apparent absorption bands are selected, indicated by red points in Figure \ref{fig:spectrum}, ignoring the narrow atomic lines listed in \citet{2016A&A...595A.105D}. Then, a linear fitting is performed to these points, and the original spectrum is divided by this fitting line to get the re-normalized spectrum. At last, the continuum of the spectrum is defined as the Gaussian mean of the points used for the linear fitting to correct for possible systematic offset. Figure \ref{fig:spectrum} shows an example of this process, with the CaT measurement region marked in green. For consistency, the re-normalization is applied to all spectra, and the spectra of late-type stars are effectively flattened, while no significant difference is observed in the spectra of other-type stars before and after the correction.

\subsection{Selection of the initial sample of Galactic stars}
\subsubsection{Data pre-processing}

To ensure that the analyzed spectra are from Galactic stars, the following criteria are applied:

\begin{enumerate}
\item A stellar probability greater than 99\% ({\tt classprob\_dsc\_combmod\_star}$>0.99$);
\item Stars not located in the LMC, i.e. the stars within the region $64^\circ<{\rm R.A.}<98^\circ$, $-78^\circ<{\rm Decl.}<-59^\circ$ and RV greater than 100 km/s, are excluded;
\item Stars not located in the SMC, i.e. the stars within the region $2^\circ<{\rm R.A.}<26^\circ$, $-76^\circ<{\rm Decl.}<-69^\circ$ and RV greater than 100 km/s, are excluded;
\end{enumerate}

As a result, 7407 spectra are excluded. From the remaining spectra, only those with a SNR greater than 100 ({\tt rvs\_spec\_sig\_to\_noise} $>100$) are chosen. This is done to ensure that the subsequent analysis is performed on high-quality spectra, minimizing the impact of noise on the equivalent width measurements. In the end, 118,048 spectra are retained for further analysis.

\subsubsection{Removal of early-type stars} \label{sec:removal_early_type}

\begin{figure*}[htp]
	\centering
     \includegraphics[width=\textwidth]{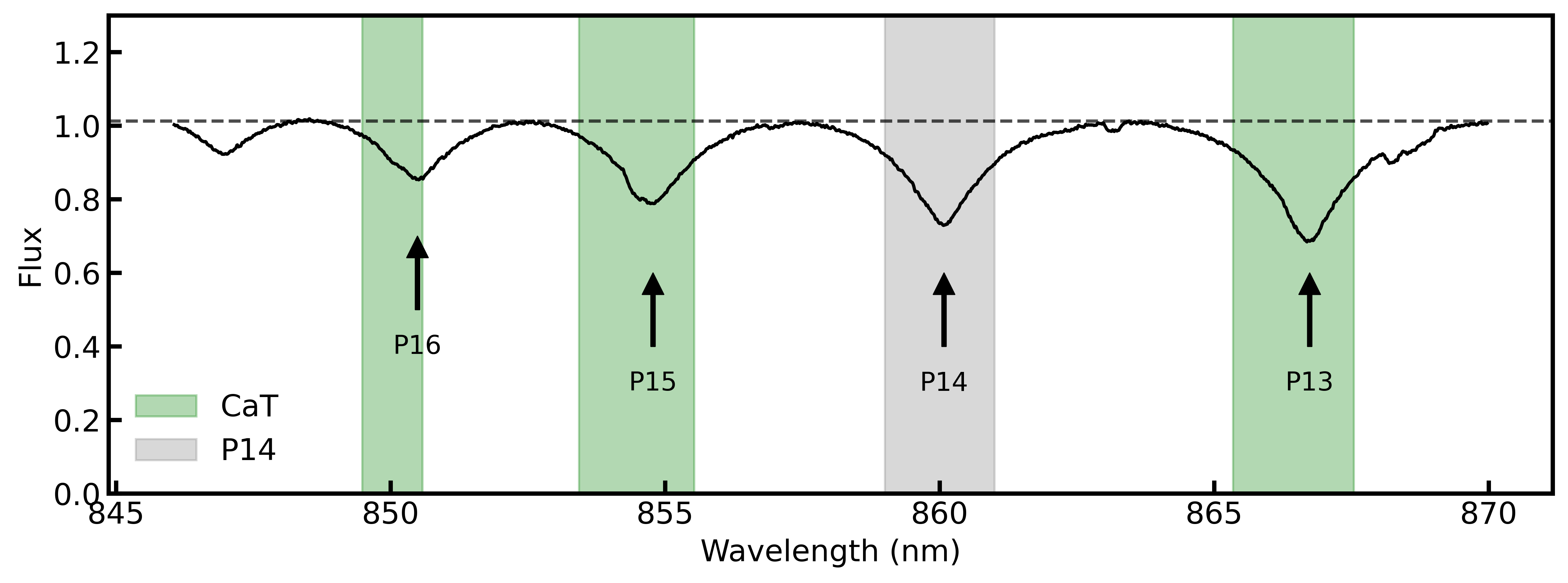}
	\caption{Example of the RVS spectrum of a typical early-type star ({\tt source\_id}: 2270570062017774976). The position indicated by the arrow marks the Paschen line series P13-P16 for hydrogen. The gray and green shade mark the measuring ranges of P14 and CaT, respectively. \label{fig:early_type_spectrum}}
\end{figure*}

Because the Paschen series lines P13, P15, and P16 in early-type stars appear at similar wavelengths as the CaT (see Figure \ref{fig:early_type_spectrum}), the measurement of the EW(CaT) may be actually conducted to the hydrogen lines. Thus it is necessary to remove the early-type stars.

Since no strong absorption lines in the spectra of late-type stars appear near P14 (see Figure \ref{fig:early_type_spectrum}), the equivalent width of P14 region (EW(P14)) is used as the diagnosis of early-type stars. EW(P14) is calculated by using the \texttt{equivalent\_width} function from the \texttt{specutils} package \citep{2019ascl.soft02012A}. The continuum is determined by using the Gaussian mean of the points for the linear fitting described in Section \ref{sec:RVS_spectra}, which is shown by the black dashed line in Figure \ref{fig:early_type_spectrum} where the continuum level is 1.013 in this case. The wavelength range is set between 859-861 nm, i.e. the shaded gray region in Figure \ref{fig:early_type_spectrum}.

\begin{figure*}
    \centering
    \includegraphics[width=0.6\textwidth]{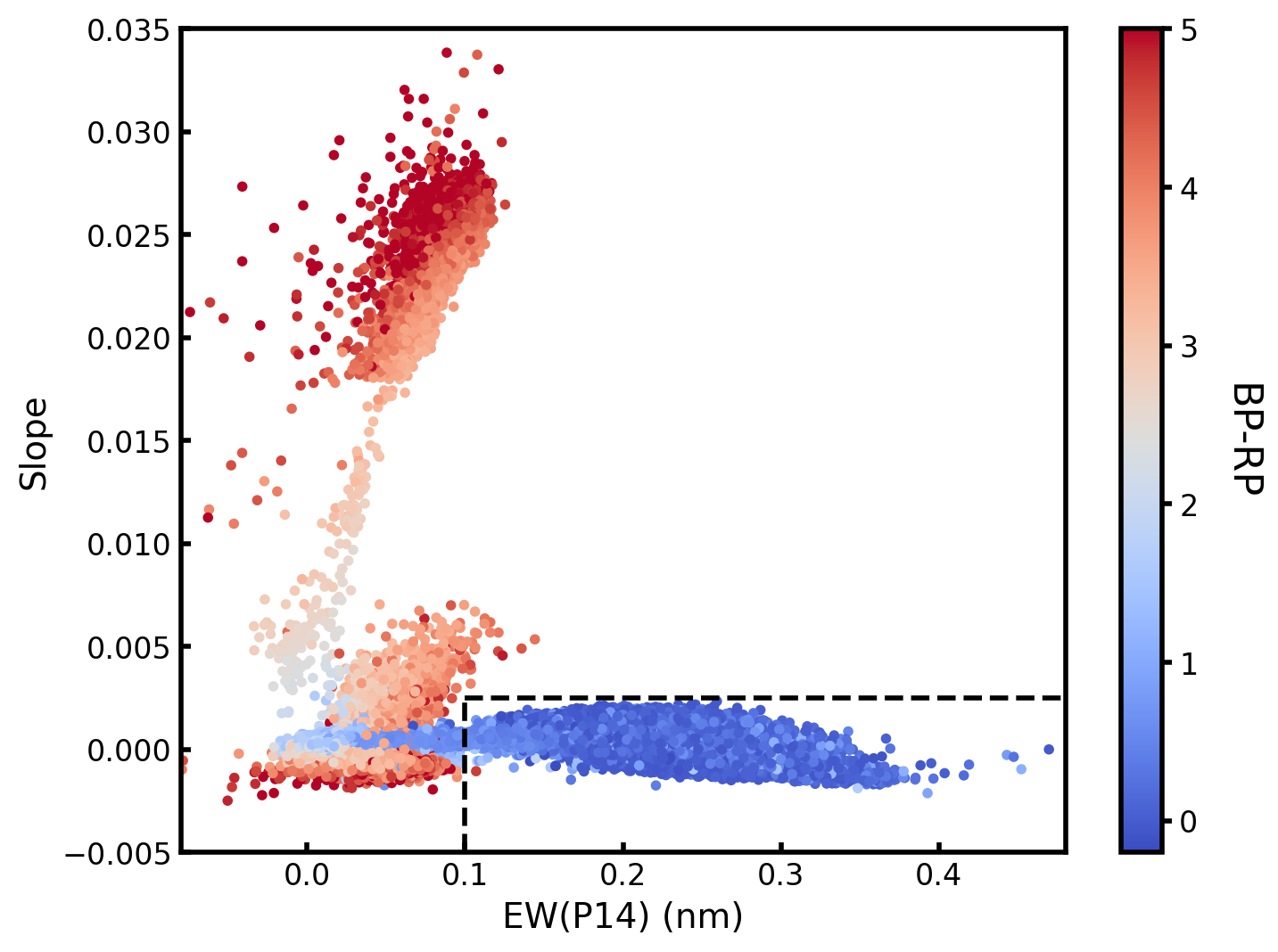}
    \caption{The diagram used to remove early-type stars. The horizontal and vertical axes are EW(P14) and the slope mentioned in Section \ref{sec:RVS_spectra}, respectively. The black dashed box defines the area of early-type stars, and the dots are color-coded by $BP-RP$. }
    \label{fig:early_type}
\end{figure*}

In additon to the EW(P14), the slope of the original Gaia RVS continuum derived from the linear fitting in Section \ref{sec:RVS_spectra} is supplemented to assure the identification of early-type stars. Because the continuum of early-type stars are not affected by broad molecular bands, the slope should be small. As shown in Figure \ref{fig:early_type}, stars with significant EW(P14) indeed exhibit a small continuum slope, which supports this criterion. Specifically, stars with EW(P14) greater than 0.1 nm and a slope less than 0.0025 are classified as early-type. Consequently, a total of 12,396 stars are excluded, which are located in the black dashed box in Figure \ref{fig:early_type}. Among the initial 118,048 stars in the sample, 8,157 have available {\tt teff\_esphs} values, indicating temperatures above 7,500 K, and are thus considered hot stars \citep{2023A&A...674A..26C}. Notably, of the 12,396 early-type stars excluded, 8,133 ($\sim$99.7\%) fall within this group, further validating the effectiveness of the early-type star identification and exclusion process.

\section{The Galactic RSGs candidates}  \label{sec:RSG}

The RSG candidates are basically selected by the EW(CaT) in that it is generally larger for supergiants than dwarfs or giants. However, the Galactic RSGs are mostly M-type \citep{2012AJ....144....2L}, leading to the decrease of EW(CaT) to be comparable to that of giants or even dwarfs \citep{2016ApJ...821..131J}. Thus other parameters including the color index $BP-RP$ and $K-W3$ and the equivalent width of TiO band are taken into account. Accordingly, the RSG candidates are divided into the early-type (K0-M2) and late-type (after M2) groups during the process.

\subsection{Preliminary sample of RSGs in the EW(CaT) v.s. $BP-RP$ diagram} \label{sec:dwarf_giant}

\subsubsection{EW(CaT) of RSGs} \label{sec:CaT_of_RSGs}

The CaT is a characteristic absorption feature in the near-infrared spectrum, with wavelengths centered at 8498, 8542, and 8662 \r{A} in air (see e.g. \citet{2021A&A...654A.130C,2023A&A...674A..26C, 2023A&A...674A..28F}). It serves as a strong indicator of luminosity class (i.e., surface gravity), where lower surface gravity typically corresponds to stronger CaT absorption, resulting in a larger EW(CaT). Therefore, using EW(CaT) to identify RSGs is a reasonable approach, and this has been applied in previous studies. \citet{1988AJ.....96.1884H} classified stars with EW(CaT) greater than 1.1 nm as supergiants when analyzing RSG spectra in M31. \citet{2011A&A...528A..59N}, focusing only on the 8542 and 8662 \r{A} lines, found that Galactic RSGs earlier than M3 satisfied EW(Ca II 8542 + Ca II 8662) $>$ 0.9 nm. Within the same luminosity class, EW(CaT) increases from F- or G-type stars until it saturates at around M2, after which it rapidly decreases with later spectral type. This decline is due to the increasing prominence of TiO in late-type stars, which erodes both the continuum and atomic lines, thus reducing the equivalent width of CaT.

\begin{figure*}[h]
	\centering
    \includegraphics[width=0.6\textwidth]{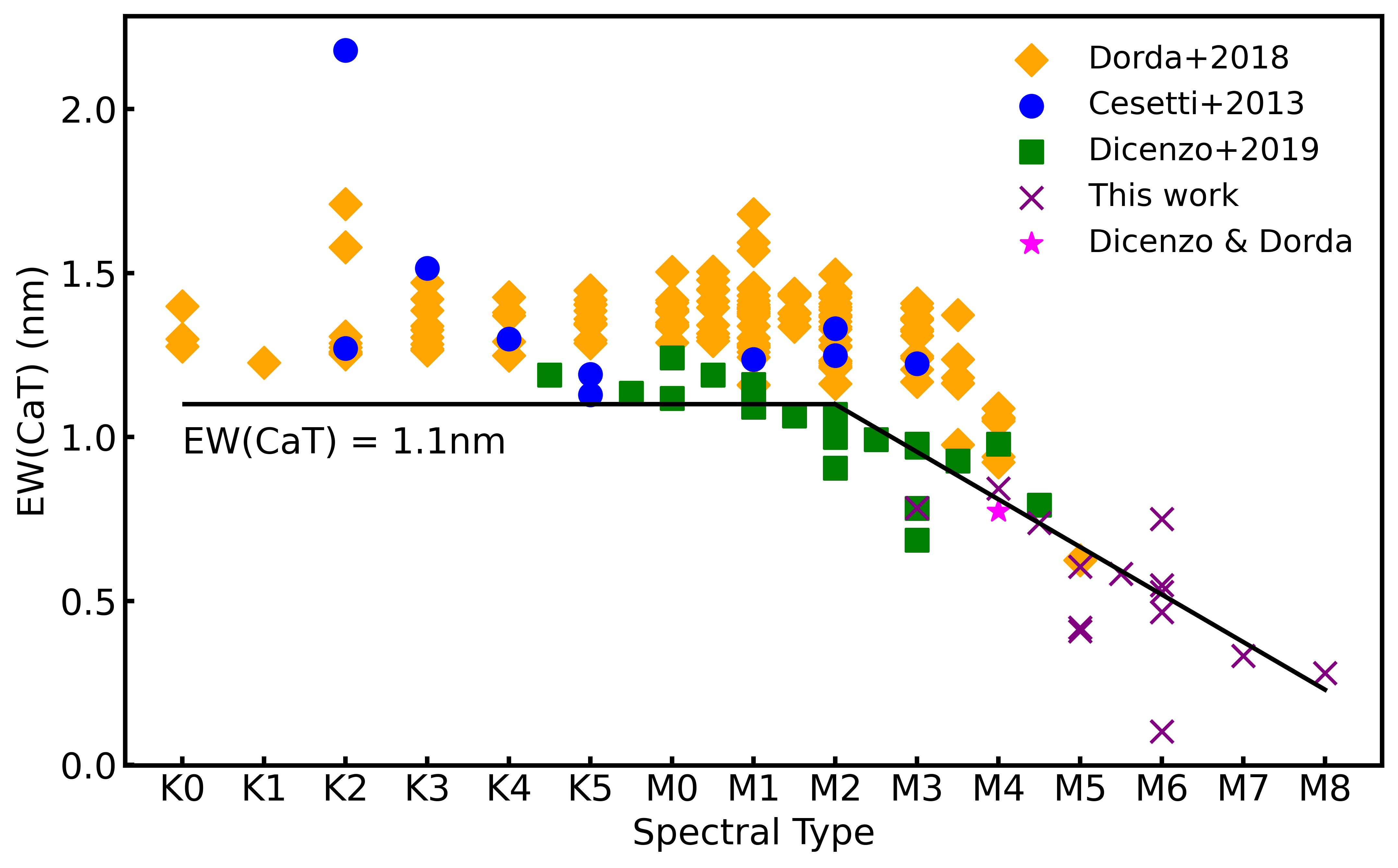}
	\caption{Spectral types versus EW(CaT) for 185 known Galactic RSGs, with their EW(CaT) measurements from \citet{2013A&A...549A.129C}, \citet{2018MNRAS.475.2003D}, \citet{2019AJ....157..167D}. For the only magenta asterisk, the results of \citet{2018MNRAS.475.2003D} and \citet{2019AJ....157..167D} are represented after being averaged. The crosses denote the sources measured in this work from the Gaia RVS spectra. The solid black lines mark EW(CaT) = 1.1 nm (for K-type or early M-type RSGs) and the decreasing of EW(CaT) with spectral type (for RSGs later than M2), respectively. \label{fig:standard_RSGs}}
\end{figure*}

Although this trend has been proved in previous studies (see Figure 4 in \citet{2011A&A...528A..59N} and Figure 9 in \citet{2016A&A...595A.105D}), 185 known Galactic RSGs are selected to quantitatively examine the variation of EW(CaT) with spectral type. Their EW(CaT) measurements are taken from \citet{2013A&A...549A.129C}, \citet{2018MNRAS.475.2003D}, \citet{2019AJ....157..167D} or this work with RVS spectra. Figure \ref{fig:standard_RSGs} illustrates the relation between EW(CaT) and spectral type for these stars, displaying the same trend as previously described. Early-type RSGs (up to M2) have EW(CaT) $>$ 1.1 nm which is consistent with the definition by \citet{1988AJ.....96.1884H}. In Figure \ref{fig:standard_RSGs}, the EW(CaT) value at M6 has a large dispersion, which is because of the difficulties in determining the spectral type of late-M RSGs, as they are variable. For instance, the star with the largest EW(CaT) at M6 is the well-known NML Cyg, whose spectral type is reported to range from M4.5 to M7.9 \citep{2017ARep...61...80S}. NML Cyg is a supergiant, potentially even a hypergiant, and its EW(CaT) is expected to be larger than that of stars with similar spectral types. Conversely, the star with the smallest EW(CaT) at M6 is V577 Cep, whose spectral type is reported to range from M6 to M8 \citep{2014yCat....1.2023S}. \citet{2016MNRAS.461.3296N} included V577 Cep in their catalog of nearby RSG candidates, although they claimed the possibility that their sample may contain red giants with luminosity class of II or III. A linear fitting is applied to build the relation of EW(CaT) with spectral type later than M2, requiring the fitting to pass through the M2 type at EW(CaT) = 1.1 nm, as shown by the black line in Figure \ref{fig:standard_RSGs}. This provides an approximate EW(CaT) values for RSGs of later spectral types, which is listed in Table \ref{tab:BPRP_CaT}. It can be noted that the EW(CaT) decreases to 0.230 nm at M8.

\subsubsection{$(BP-RP)_0$ of RSGs} \label{sec:color_of_RSGs}

The intrinsic color index $(BP-RP)_0$ is calculated to constrain the minimal observational $BP-RP$ for specific sub-type which is associated with the EW(CaT). In addition, RSGs are usually more luminous than other type of stars and experience higher interstellar extinction to appear redder. The MIST model \citep{2011ApJS..192....3P, 2013ApJS..208....4P, 2015ApJS..220...15P, 2016ApJS..222....8D, 2016ApJ...823..102C} is used to generate evolutionary tracks for stars with initial mass between $8-40 M_{\odot}$, assuming $v/v_{crit} = 0.4$, [Fe/H]=0, and $A_{V}=0$. Subsequently, the temperature scale for Galactic RSGs from \citet{2005ApJ...628..973L} is adopted to determine the effective temperature ($T_{\rm eff}$) for each spectral subtype. The $T_{\rm eff}$ for K1, M0, M1, M2, M3, M4, and M5 RSGs is set at 4100K, 3790K, 3745K, 3660K, 3605K, 3535K and 3450K respectively, and the values of $(BP-RP)_0$ are presented in Table \ref{tab:BPRP_CaT}.



\subsubsection{The EW(CaT) vs. $BP-RP$ diagram} \label{sec:EW_vs_color}

The EW(CaT) vs. $BP-RP$ diagram serves as a substitute of the EW(CaT) vs. spectral type diagram (Figure \ref{fig:standard_RSGs}) to define the region of RSGs. Because no spectral type information is available for all selected RVS sources, the observational $BP-RP$ is used as an indicator. The EW(CaT) of the remaining sources are calculated through the method described in Section \ref{sec:removal_early_type}, and the comparison of EW(CaT) with \citet{2018MNRAS.475.2003D} indicates high consistency, as it can be fitted with $\mathrm{EW(CaT)}_{\mathrm{Dorda+18}} = 1.1 \times  \mathrm{EW(CaT)}_{\mathrm{Thiswork}} - 0.05 \ $nm with a root-mean-square dispersion of 0.057nm. The error in equivalent width is calculated by using the equation in \citet{2006AN....327..862V}:
\begin{equation}
    \sigma(W_{\lambda})=\sqrt{1+\frac{F_\mathrm{C}}{\overline{F} }}\cdot \frac{(\Delta \lambda-W_{\lambda})}{S/N} \label{eq:error}
\end{equation}
where $F_\mathrm{C}$ represents the continuum flux, $\overline{F}$ is the average flux over the measured wavelength range, $\Delta \lambda$ is the wavelength range, $W_{\lambda}$ is the equivalent width of the line, and $S/N$ is the SNR of the spectrum. The absolute error of EW(CaT) is presented as $\sigma(\mathrm{EW(CaT)})$, and the relative error is then defined as $\sigma(\mathrm{EW(CaT)})/\mathrm{EW(CaT)}$. For example, with SNR $=$ 100 and EW(CaT) $=$ 0.5 nm, a typical $\sigma(\mathrm{EW(CaT)})$ is around 0.07 nm. According to Equation \ref{eq:error}, $\sigma(\mathrm{EW(CaT)})$ is expected to be inversely proportional to both the EW(CaT) itself and the SNR. Therefore, smaller equivalent widths and lower SNR will naturally lead to larger measurement errors.

\begin{figure*}[h]
	\centering
    \includegraphics[width=\textwidth]{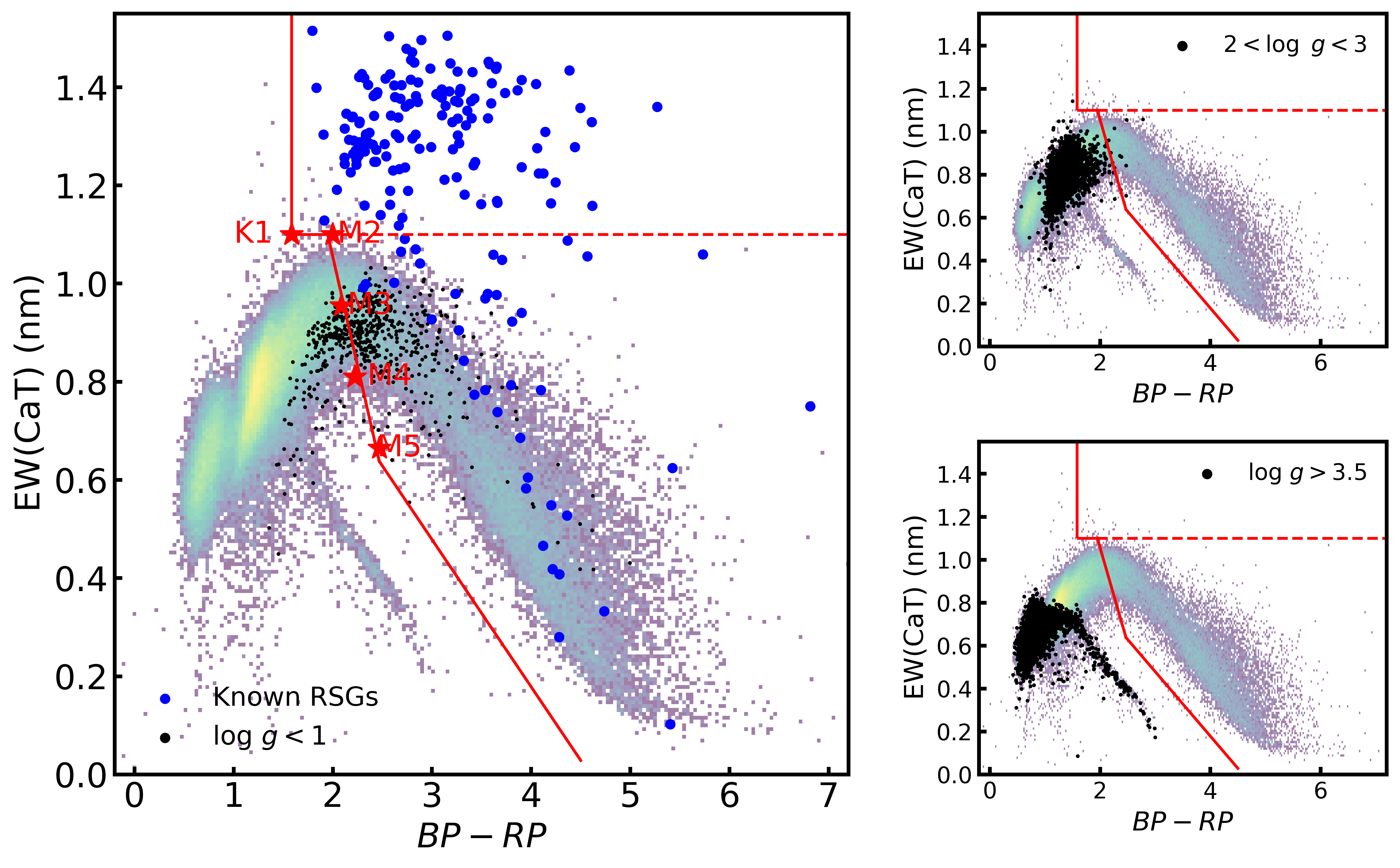}
	\caption{The diagram of EW(CaT) vs. $BP-RP$ with color coded by density for all selected RVS objects. Overlaid black dots are those with APOGEE measurements of $\log g$. They are plotted separately to represent the positions of different types of stars. The blue dots on the left panel are known Galactic RSGs mentioned in Section \ref{sec:CaT_of_RSGs}. The red asterisks mark the positions of zero-extinction K1 and M2-M5 RSGs. The red solid lines represent the bluest boundary of Galactic RSGs in this diagram, while the red dashed line marks the dividing line between the early and late type RSGs. \label{fig:CaT}}
\end{figure*}

Figure \ref{fig:CaT} presents the EW(CaT) vs. $BP-RP$ diagram, where all selected RVS sources are displayed with density-based color coding. In order to clarify the locations of different classes of stars, the sample stars are cross-identified with the APOGEE DR17 \citep{2022ApJS..259...35A} catalog which provide stellar atmospheric parameters from high-resolution near-infrared spectroscopic survey. This results in 6905 and 4295 stars with $\log\ g = 2-3$ and $\log\ g > 3.5$, which are considered as red giants and dwarfs, respectively. As displayed in the right panel of Figure \ref{fig:CaT}, the area of the dwarfs and red giants are clearly defined, which are two high-density regions with EW(CaT) equal to 0.4-0.7 nm and 0.6-0.9 nm, respectively. Additionally, the branch that decline from 1.5 to 3 in $BP-RP$ is of M-dwarfs, which is confirmed by their high proper motion measured by Gaia, because they are close to us.

With the results from Section \ref{sec:CaT_of_RSGs} and \ref{sec:color_of_RSGs}, the bluest boundary of the Galactic RSGs is delineated in Figure \ref{fig:CaT} by the solid red line. The upper lines are determined by the $(BP-RP)_0 = 1.548$ for K1-type RSGs and an EW(CaT) value of 1.1 nm before M2. For M2 to M5 types, a linear fitting between EW(CaT) and $(BP-RP)_0$ is used to describe the approximate bluest boundary of RSGs in the EW(CaT) vs. $BP-RP$ diagram. Since the $T_{\rm eff}$ and $(BP-RP)_0$ for RSGs later than M5 are unknown, a roughly parallel line to the right branch of the RVS sources is manually drawn to indicate the boundary for later-type RSGs. The known Galactic RSGs (as described in Section \ref{sec:CaT_of_RSGs}) indicated by blue dots in Figure \ref{fig:CaT} are all on the right side of the boundary lines. Moreover, 576 stars with $\log\ g < 1$ from the APOGEE catalog, which are considered to be candidates of supergiants, are also located in this area. Further inspection reveals that sources located to the left of the boundary line have lower metallicity, indicating that they are unlikely to be young, metal-rich RSGs. Both results demonstrate that the boundary lines are reasonable.

Following the boundary lines, the RSG candidates are distributed in two areas in Figure \ref{fig:CaT}, which are separated by the red dashed line at EW(CaT) $=$ 1.1 nm. Stars above and below this line are classified into the early-type and late-type RSG candidates, respectively, because the former have EW(CaT)$>$1.1 nm, and the latter have EW(CaT) less than this value. As to the late-type candidates, the EW(CaT) decreases with $BP-RP$, suggesting these stars are late-M type bright stars. Although the intrinsic color index of RSGs is almost the same as dwarfs or giants, the observed color index is significantly red because high-luminosity RSGs experience much more extinction. Therefore these objects are on the right branch in Figure \ref{fig:CaT} as expected.

\subsection{The early-type RSG candidates} \label{sec:early_type_RSGs}

As described in Section \ref{sec:EW_vs_color}, early-type RSG candidates in our sample are defined as sources with $BP-RP > 1.584$ and EW(CaT) $>$ 1.1 nm. Out of the remaining 105,652 sources, 30 meet these criteria, which are presented in Table \ref{tab:early_RSGs}. The relative errors of EW(CaT) of all 30 stars are below 6\%, as their EW(CaT) values are sufficiently large, ensuring precise measurements. Gaia DR3 provides stellar spectral types through the column {\tt spectraltype\_esphs} in the astrophysical\_parameters table, which includes OBAFGKM and C stars. This classification is primarily based on specific spectral features in BP/RP spectra, such as the dominance of CN and $C_{2}$ bands in carbon stars, and the presence of TiO and VO in M-type stars \citep{2023A&A...674A..15L, 2023A&A...671A.148M}. Among the 30 early-type RSG candidates, 24 and 4 are marked as K-type and M-type stars, respectively, which aligns with expectations since these candidates are expected to be late-K or early-M types. One source is marked as a Cstar ({\tt source\_id}: 5853442496285943424). But its XP spectrum does not show any significant carbon absorption bands when compared with the XP spectra library of C-rich star of \citet{2023A&A...671A.148M}, suggesting this is likely a misclassification by Gaia. Another source is marked as an O-type star ({\tt source\_id}: 5546711192039738752), but its XP spectrum is not blue-peaked, while with clear TiO absorption features, indicating this is a late-type M star.

Regarding stellar parameters, five stars in the sample have available Apsis GSP-phot parameters, which are derived by fitting XP spectra using PHOENIX or MARCS models to provide best-fit estimates of $T_{\mathrm{eff}}$, $\log\ g$, and [Fe/H] \citep{2023A&A...674A..27A, 2023A&A...674A..28F}. Their $T_{\mathrm{eff}}$ values are 4451K, 4455K, 4524K, 4568K, and 5123K, with $\log\ g$ values between 0.87 and 1.58, consistent with early-type RSG characteristics. However, as described in \citet{2023A&A...674A..28F}, the {\tt logposterior\_gspphot} column indicates the quality of the Apsis GSP-phot fitting, with higher values for better fittings. It is recommended to only use results with {\tt logposterior\_gspphot} $> -1000$. Unfortunately, the highest value among these five stars is -9000. As \citet{2023A&A...671A.148M} pointed out, the Gaia Apsis pipeline struggles to handle bright late-type stars due to their significant variability and fitting challenges, suggesting that these parameters may not be highly reliable and should be treated cautiously.

Additionally, 28 stars in the sample have Apsis GSP-spec parameters, which are derived from RVS spectra \citep{2023A&A...674A..29R}. Following recommendations from \citet{2023A&A...674A..29R} and \citet{2023A&A...671A.148M}, sources are filtered by requiring the first nine digits of {\tt flags\_gspspec} to be less than or equal to 1, and the 10th to 12th digits to be non-null. Moreover, \citet{2023A&A...674A..29R} indicated that due to parameterization issues in GSP-Spec, cool stars with $T_{\mathrm{eff}} \lesssim 4000\mathrm{K}$ had their parameters set to $T_{\mathrm{eff}}=4250\pm 500\mathrm{K}$ and $\log\ g=1.5\pm 1$. After accounting for these limitations, eight stars remain with reliable parameters. Their $T_{\mathrm{eff}}$ ranges from 3782K to 4632K, and $\log\ g$ values span from -0.29 to 0.65, which are also consistent with typical RSGs.

In summary, the stellar parameters derived from the Gaia Apsis pipeline coincides with RSGs for the 30 early-type candidates, in spite that they suffer large uncertainty.

\subsection{The late-type RSG candidates}
\subsubsection{TiO band} \label{sec:TiO}

As described in Section \ref{sec:EW_vs_color}, late-type RSGs can only be distributed along the right branch of Figure \ref{fig:CaT}, where EW(CaT) $<$ 1.1 nm, and on the right side of the red solid line. This subset contains 20,873 stars. Because late-type RSGs are cool, oxygen-rich stars, their spectra should feature strong TiO bands. Typically, TiO absorption starts to appear in M2-type stars, peaks at M6, and then saturates. Prominent TiO bands in RSGs include those at 8432+8442+8452 \r{A} and 8859 \r{A}, unfortunately, none of these are covered by the RVS spectra. On the other hand, the XP spectra covers a wide wavelength range, including several strong TiO bands, which are used to screen oxygen-rich M-type stars.

\begin{figure*}[h]
	\centering
    \includegraphics[width=0.6\textwidth]{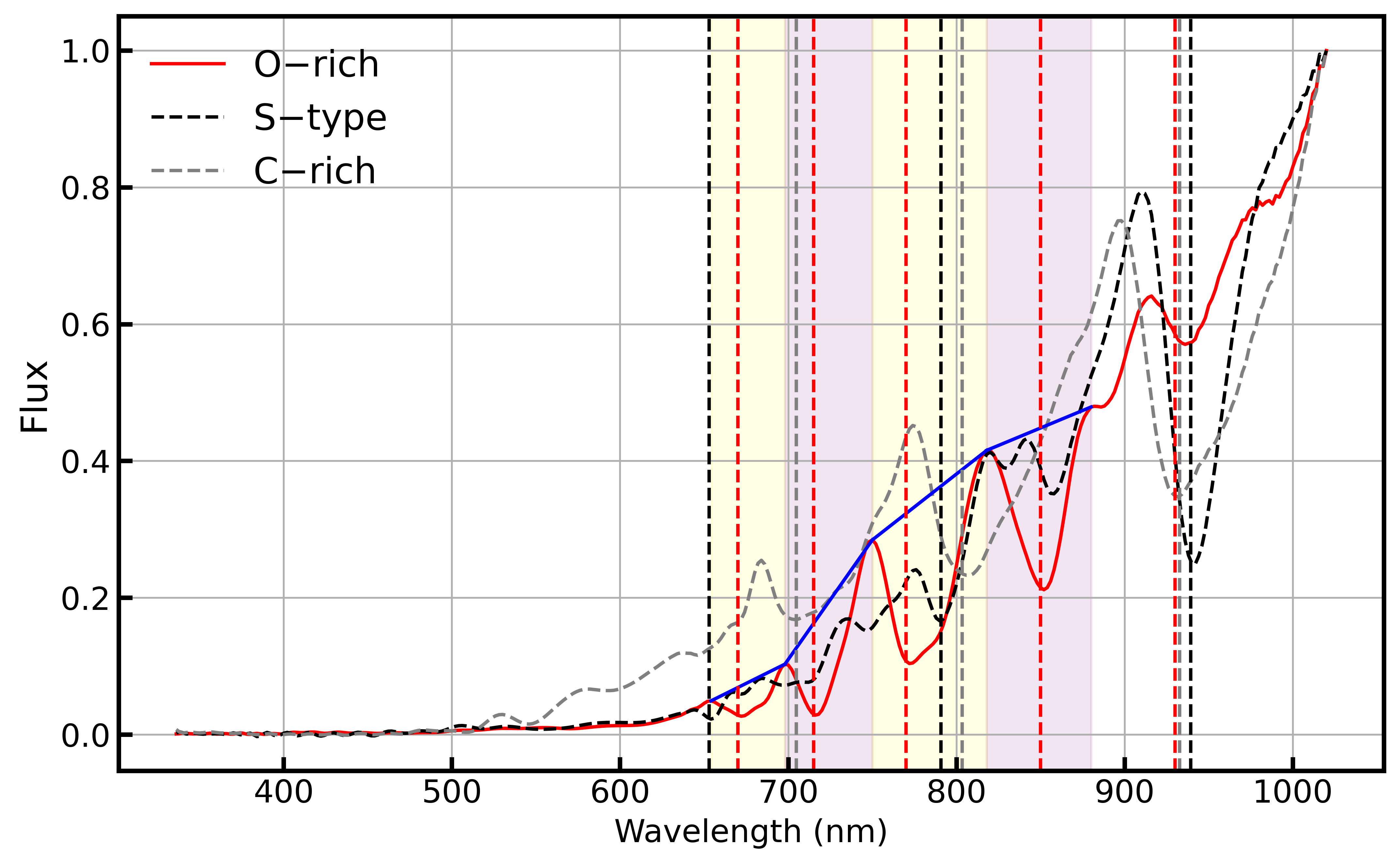}
	\caption{Examples of three XP spectra of an oxygen-rich star ({\tt source\_id}: 5853243278519978112, red solid line), a carbon-rich star ({\tt source\_id}: 5716487091710504064, gray dashed line), and an S-type star ({\tt source\_id}: 5233074194539855360, black dashed line). The spectra are normalized to the maximum flux for comparison. The shaded color is the ranges of the four TiO that is measured. The four blue solid lines represent the pseudo-continuum of measuring equivalent width.\label{fig:XP_example}}
\end{figure*}

\citet{2023A&A...671A.148M} identified five TiO absorption features in their XP spectral library, centered approximately at 670, 715, 770, 850, and 940 nm. Given that the 940 nm region also includes absorption from ZrO and CN, only the first four bands are used. Specifically, four wavelength ranges are examined, i.e. 654-698 nm, 698-750 nm, 750-818 nm, and 818-880 nm. If a single local minimum is detected within any of these ranges, TiO absorption is considered present, and its equivalent width is measured. The equivalent width is calculated by linearly fitting the endpoints of each range to represent a pseudo-continuum, following the method described in Section \ref{sec:removal_early_type}. This approach is reasonable because the endpoints for each range are chosen at the flux maxima of true M-type oxygen-rich stars, and the region between two adjacent flux peaks is designated as the TiO absorption band as shown in Figure \ref{fig:XP_example}. This also minimizes the potential contamination by C or S-type stars, as their dominant molecular bands and flux minima are located at different wavelengths, producing distinctly different spectral shapes \citep{2023A&A...674A..15L}. It can be seen in Figure \ref{fig:XP_example} that the spectral shape of C star and S-type star is quite different from oxygen-rich star, making the classification reliable. For insurance, only sources that exhibit TiO absorption in at least two of the four bands are considered true M-type oxygen-rich stars. The equivalent widths of all detectable TiO bands are averaged to yield a single EW(TiO) value. Consequently, 1,480 stars are excluded from the sample, 848 of which are excluded due to the lack of XP spectra.

\begin{figure*}[h]
	\centering
    \includegraphics[width=0.6\textwidth]{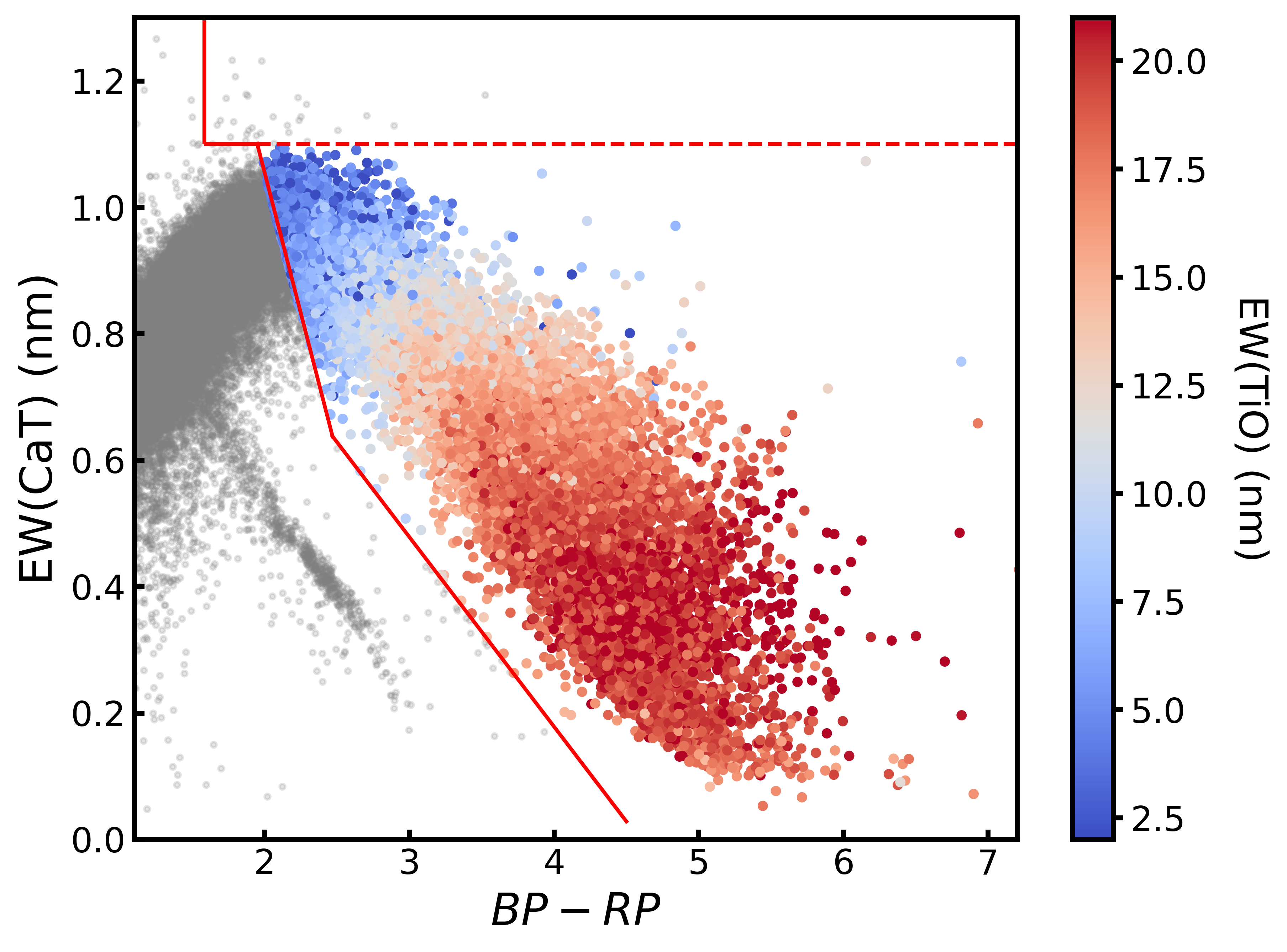}
	\caption{The same as Figure \ref{fig:CaT}, but the regions for late-type RSG candidates are color-coded with EW(TiO), and the rest are shown by gray dots. \label{fig:EW_TiO_branch}}
\end{figure*}

\begin{figure*}[h]
	\centering
    \includegraphics[width=\textwidth]{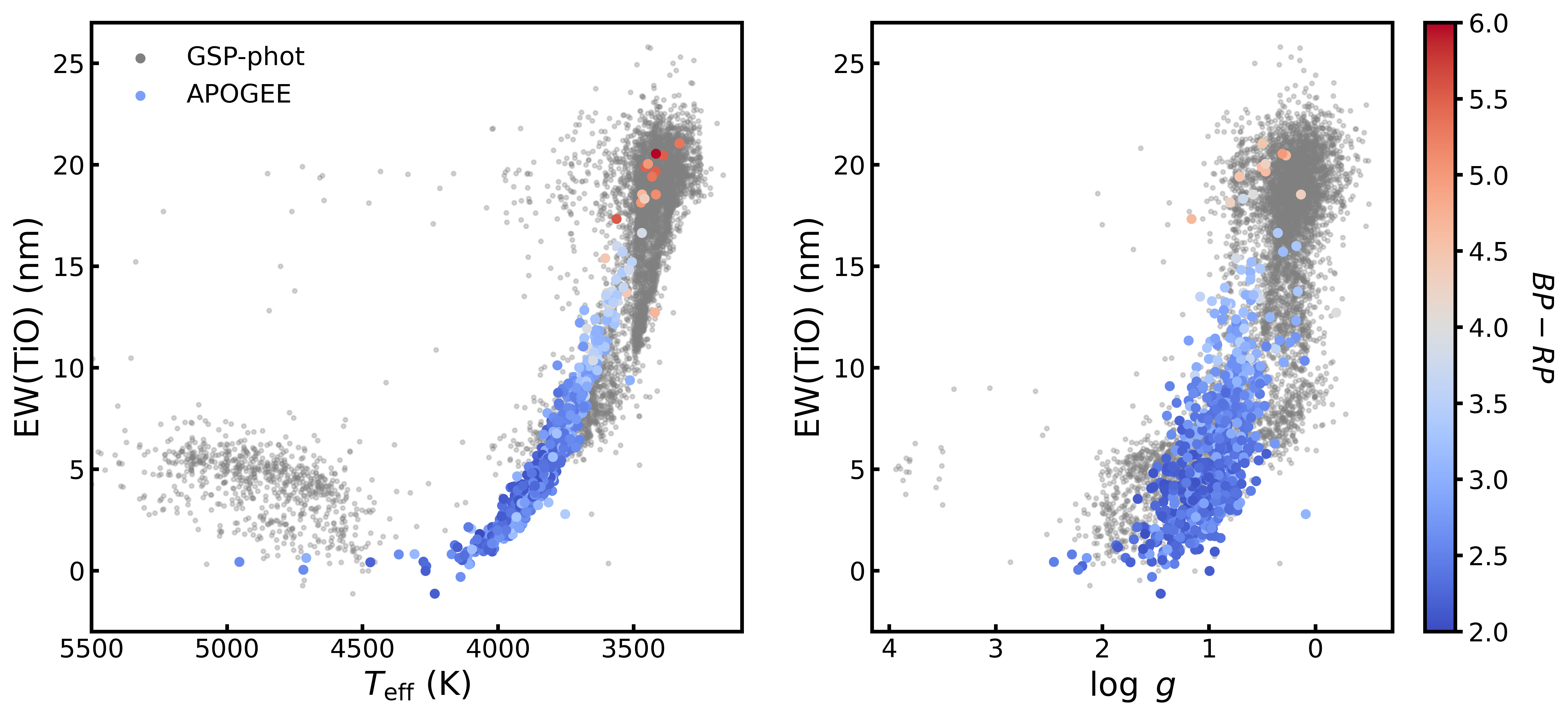}
	\caption{The relation of EW(TiO) with $T_{\mathrm{eff}}$ and $\log\ g$. The dots color-coded by $BP-RP$ are APOGEE measurements, while the gray dots are from Gaia GSP-phot parameters. There are 655 and 5934 stars have available APOGEE and GSP-phot parameters, respectively. \label{fig:EW_TiO_teff_logg}}
\end{figure*}

Figure \ref{fig:EW_TiO_branch} shows the distribution of EW(TiO) in the right branch of EW(CaT) vs. $BP-RP$ diagram. It reveals that smaller and larger EW(TiO) values correspond to bluer and redder sources, respectively. Since TiO is a reliable indicator of spectral type, the descending branch on the right side of Figure \ref{fig:EW_TiO_branch} corresponds to an increase in spectral type. This is further supported by the relation between EW(TiO) and $T_{\mathrm{eff}}$ from APOGEE measurements, as shown in the left panel of Figure \ref{fig:EW_TiO_teff_logg}. A comparison between EW(TiO) and $\log\ g$ from APOGEE is also presented in the right panel of Figure \ref{fig:EW_TiO_teff_logg}, which shows that higher EW(TiO) values correspond to lower $\log\ g$, and vice versa, consistent with the conclusions of \citet{2011A&A...528A..59N} (see their Figure 3). The GSP-phot parameters are used to check the relation because they are available for many more sources, although with relatively large uncertainty. As shown in gray dots in Figure \ref{fig:EW_TiO_teff_logg}, this trend is confirmed.

\begin{figure*}[h]
	\centering
    \includegraphics[width=0.6\textwidth]{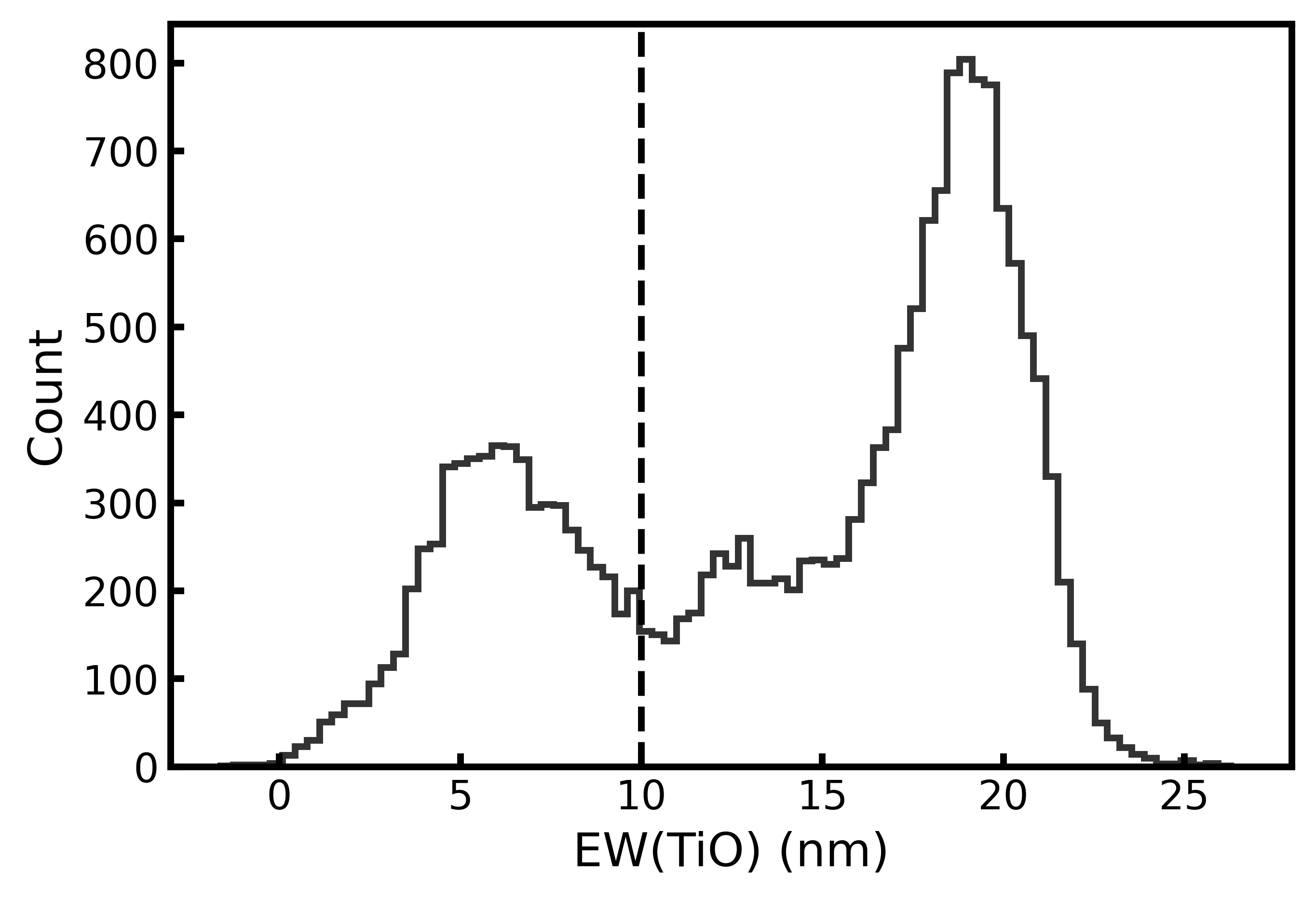}
	\caption{The histogram of EW(TiO). The black vertical line marks the position where EW(TiO) = 10 nm.\label{fig:EW_TiO}}
\end{figure*}

The remaining 20,873 stars up to present include late-type RGBs, RSGs, and AGB stars. There are two reasons to suggest that the blue dots in Figure \ref{fig:EW_TiO_branch} likely represent late-type RGB stars, i.e., close to the tip-RGB. First, \citet{2023MNRAS.523.2283D} identified metal-poor tip-RGB stars at high Galactic latitudes, with $BP-RP$ ranging from 1.55 to 2.25. For sources with higher metallicity or lower Galactic latitudes, the color would be redder, which aligns with the stars with $BP-RP < 3$ (blue dots in Figure \ref{fig:EW_TiO_branch}). Second, these blue points generally exhibit EW(TiO) values below 5-10 nm, which correspond to sources with $\log\ g > 1$ in Figure \ref{fig:EW_TiO_teff_logg}, a characteristic of bright giants. Additionally, Figure \ref{fig:EW_TiO} shows a bimodal distribution of EW(TiO), further indicating the presence of two distinct stellar populations. To minimize contamination from late-type red giants, only sources with EW(TiO) $>$ 10 nm are selected, leaving 13,316 stars.

\subsubsection{VO band}
Both TiO and VO are excellent temperature indicators. After TiO, VO only starts to appear in extremely late-type stars, beginning with M6-type \citep{1991ApJS...77..417K, 1992AJ....104..821M}. In Section \ref{sec:TiO}, it is mentioned that the descending branch on the right side of Figure \ref{fig:CaT} should indicate an increase in spectral type. To further support this point, the only VO band within the RVS spectral range at 8624 \r{A} in air, \citep{2016A&A...595A.105D} is measured.

\begin{figure*}[h]
	\centering
    \includegraphics[width=0.6\textwidth]{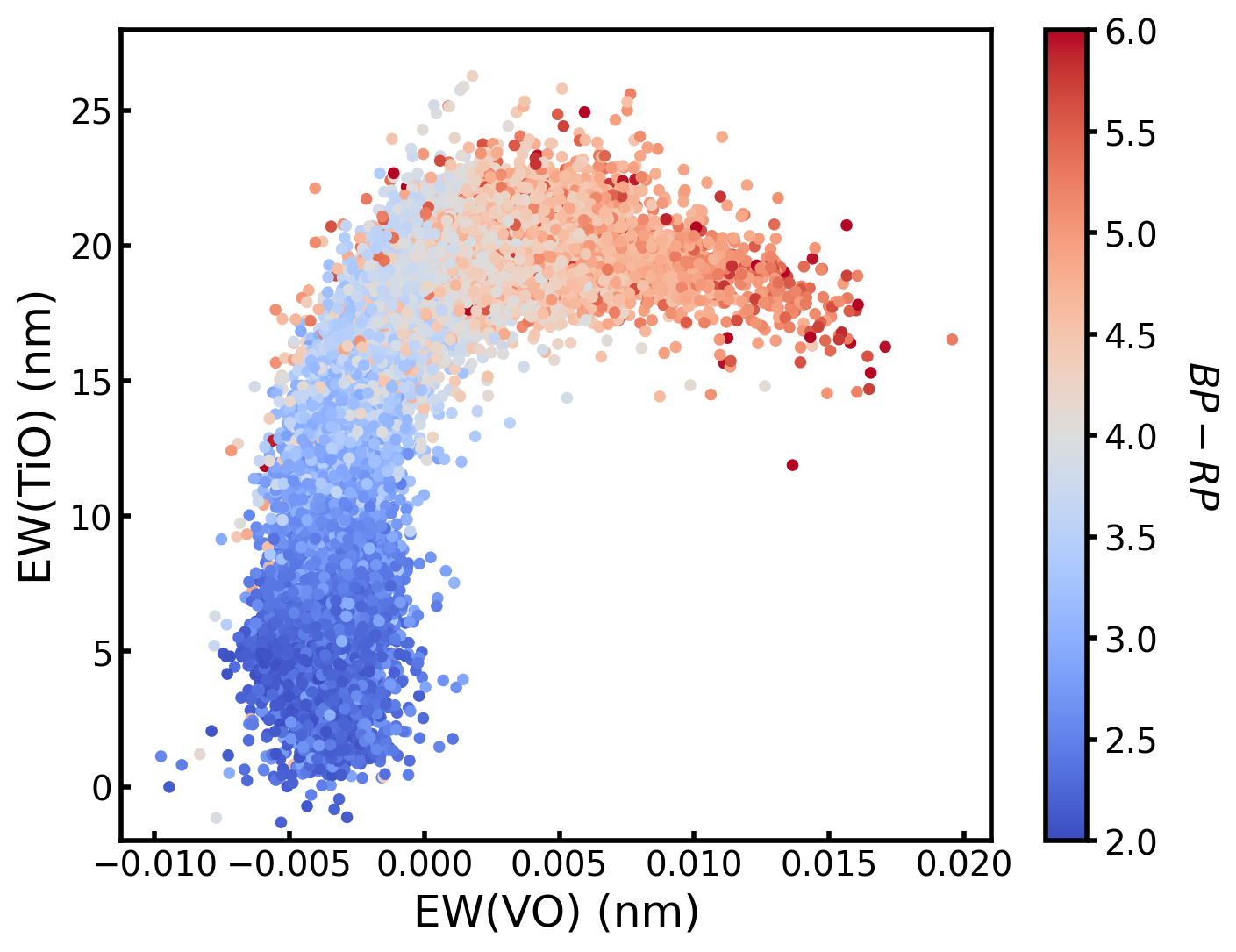}
	\caption{The EW(TiO) vs. EW(VO) diagram for all sources exhibiting TiO absorption. The dots are color coded by $BP-RP$.\label{fig:TiO_VO}}
\end{figure*}

For this VO band, following the analysis of \citet{2016A&A...595A.105D}, its equivalent width is measured within the range 8624.25-8625.5 \r{A} in air, corresponding to 862.66-862.79 nm in vacuum, and the value is denoted as EW(VO). Figure \ref{fig:TiO_VO} presents the relation between EW(VO) and EW(TiO), showing all the sources with TiO absorption (i.e., including the sources excluded in Section \ref{sec:TiO} with EW(TiO) < 10 nm). The intensity of the TiO band increases with spectral type (i.e., as temperature decreases), which manifests as a vertical branch on the left side of the figure. TiO absorption saturates around M6, at which point VO begins to appear and strengthens, forming a horizontal branch on the right side. It is evident that the horizontal branch corresponds to redder sources indicated by a large $BP-RP$, which aligns with the far-right region in Figure \ref{fig:CaT}. This suggests that the latest-type stars are located in this region. However, since the red wing of the diffuse interstellar band (DIB) near 8621 \r{A} may invade the VO band \citep{2021A&A...645A..14Z, 2023A&A...674A..40G, 2024A&A...683A.199Z}, and this VO band is so weak that it is difficult to measure at low SNR, this analysis is only presented for demonstration and is not used as a selection criterion.

\subsubsection{Removal of AGB stars}

At this point, the only potential contamination in the sample comes from AGB stars. Given that AGB stars are generally long-period variables (LPV) with more significant mass loss and larger variation amplitudes than RSGs, the diagram of $K-W3$ vs. $BP$ amplitude and period-amplitude sequence are used to exclude mass-losing AGB stars and low mass-loss AGB stars, respectively.

\begin{figure*}[h]
	\centering
    \includegraphics[width=\textwidth]{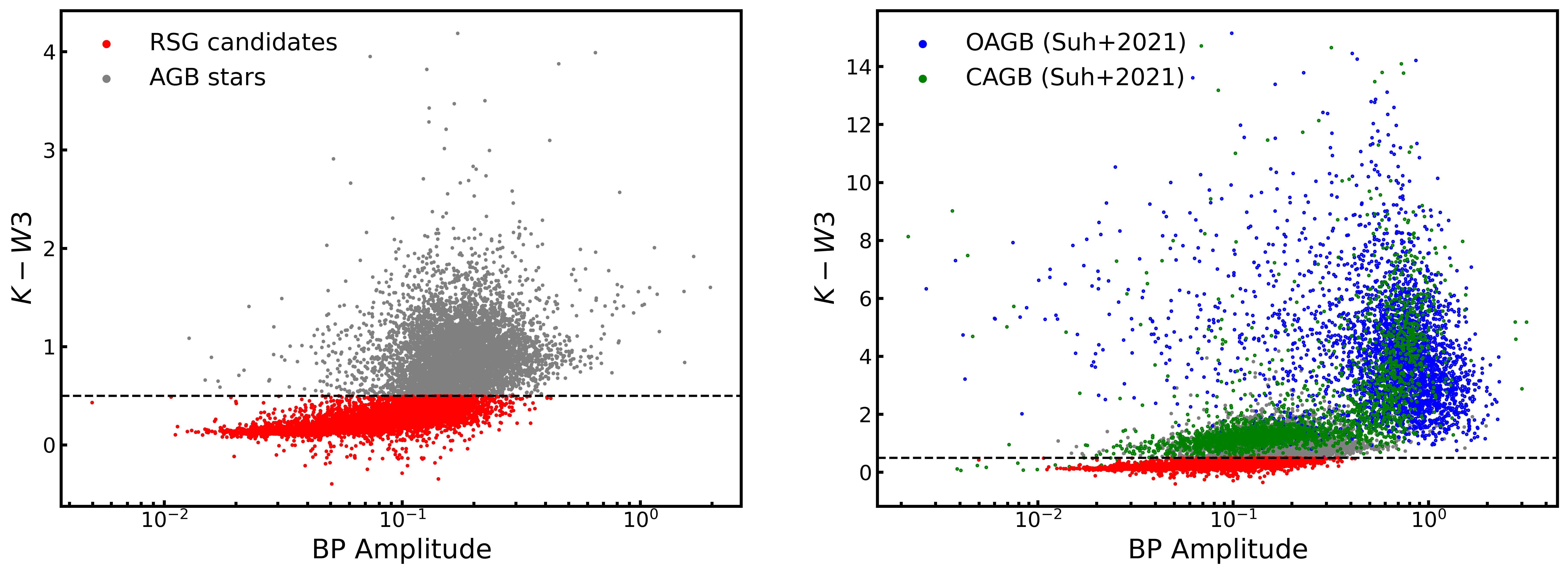}
	\caption{The $K-W3$ vs. $BP$ amplitude diagram for stars that have EW(TiO) $>$ 10 nm. The $BP$ amplitude is defined by Equation \ref{eq:BPamp}. The red dots in both panels are selected late-type RSG candidates, while the gray dots are stars that are considered to be AGB stars. The blue and green dots in the right panel are OAGB and CAGB identified by \citet{2021ApJS..256...43S}, respectively. \label{fig:KW3_BPamp}}
\end{figure*}

The $K$ and $W3$ magnitudes are taken from the Two Micron All Sky Survey (2MASS, \citet{2006AJ....131.1163S}) and ALLWISE \citep{2014yCat.2328....0C} catalog respectively. The $BP$ amplitude is defined by \citet{2017MNRAS.466.4711B} as
\begin{equation}
    BP\ \mathrm{Amplitude} =  \sqrt{{\tt phot\_bp\_n\_obs}} \times \frac{{\tt phot\_bp\_mean\_flux\_error}}{{\tt phot\_bp\_mean\_flux}}. \label{eq:BPamp}
\end{equation}
All the remaining sources are shown in Figure \ref{fig:KW3_BPamp} (left panel), together with the Galactic AGB stars identified by \citet{2021ApJS..256...43S} using IRAS data (right panel). Considering the color index $K-W3$ is a well-known indicator sensitive to dust emission \citep{2016ApJS..224...23X}, and the AGB stars from \citet{2021ApJS..256...43S} mostly have $K-W3 > 0.5$, such sources in our sample are identified as AGB stars and removed. This criterion is reinforced by the fact that the excluded sources also exhibit larger amplitudes. This process removed 6533 stars from our sample. It is important to note that $\sim$1400 stars in the $K$ band and $\sim$100 stars in the $W3$ band in the sample have poor photometric quality, with their flags not marked as 'A' in the catalog. Further examination reveals that these sources are the brightest stars in the sample, with nearly all having $K$-band magnitude $<$ 4.5 and $W3$-band magnitude $<$ 0.5, respectively. Given that such bright stars are more likely to be RSGs, no restriction is applied to the photometric quality for this subset. Instead, only stars with $K-W3 > 0.5$ are excluded, as they are more likely to be AGB stars.

\begin{figure*}[h]
	\centering
    \includegraphics[width=0.6\textwidth]{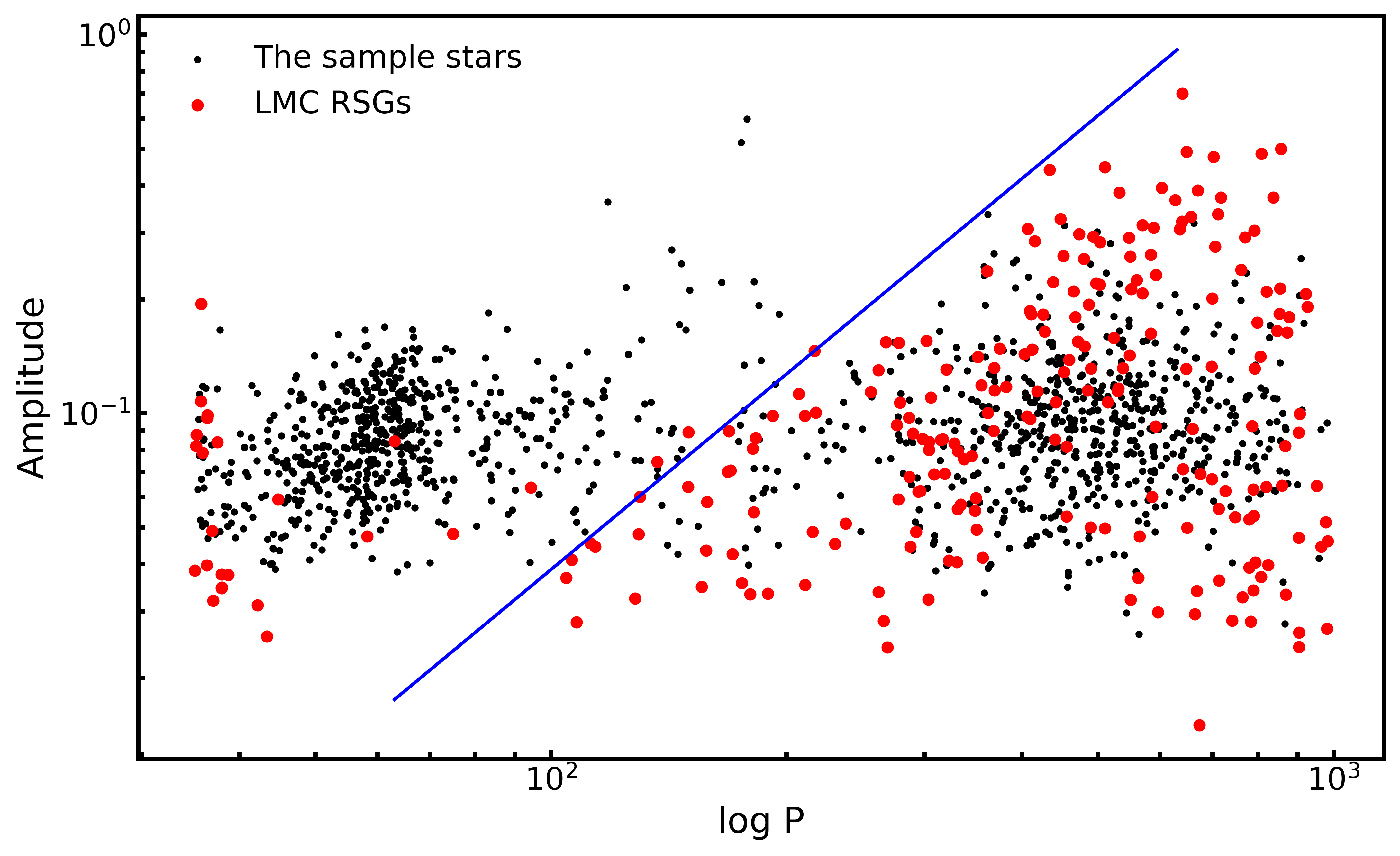}
	\caption{The period-amplitude diagram of our sample stars (black dots) and the LMC RSGs from \citet{2021ApJ...923..232R} (red dots). The amplitude and period are taken from Gaia DR3 LPV catalog \citep{2023A&A...674A..15L}. The blue solid line are manually drawn to separate the AGB stars branch (short period) and the RSGs candidate branch (long period). \label{fig:PAD}}
\end{figure*}

Low mass-loss AGB stars can be identified by examining their period-amplitude sequences, for which the LPV catalog in Gaia DR3 \citep{2023A&A...674A..15L} provides an excellent opportunity. Observations of the LMC have confirmed that LPVs build at least five sequences on the period-luminosity diagram (PLD): A, B, C', C, and D (see e.g., \citet{1999IAUS..191..151W}). Sequences A and B have been further divided into several sub-sequences to characterize small-amplitude red giant variables \citep{2007AcA....57..201S}. The PLD is a powerful tool for tracing the evolution of LPVs, as different sequences are often associated with varying metallicity and mass-loss rates \citep{2012ApJ...753...71R}. \citet{2019MNRAS.484.4678M} studied the characteristics of mass-loss rates on the period-amplitude sequences, finding that sequences A and B typically consist of AGB stars with small amplitudes and low mass-loss rates (less than $10^{-9} M_{\odot}\ yr^{-1}$). On the other hand, \citet{2024IAUS..376..292J} demonstrated that RSGs tend to fall on sequences a2 and C in the PLD. Given the well-studied PLD of the LMC, the RSGs in the LMC can be used to mark a reference region on the period-amplitude diagram, with sequences to the left of the region likely corresponding to low mass-loss AGB stars. By cross-matching the remaining sample with the Gaia DR3 LPV catalog \citep{2023A&A...674A..15L}, there are 1,180 stars with available amplitude and period. Their distributions on the period-amplitude diagram are shown as black points in Figure \ref{fig:PAD}. For comparison, 235 LMC RSGs from the \citet{2021ApJ...923..232R} sample, cross-matched with the Gaia DR3 LPV catalog, are displayed as red dots. As shown in Figure \ref{fig:PAD}, the black dots divide into two branches corresponding to shorter and longer periods, while RSGs generally exhibit periods longer than 100 days. This make sense because RSGs with pulsation periods shorter than 100 days are rarely detected \citep{2019ApJS..241...35R,2019MNRAS.487.4832C}. RSGs with timescales shorter than 100 days usually exhibit irregular variation \citep{2020ApJ...898...24R,2024ApJ...969...81Z}, making it challenging to detect in Gaia's long-cadence light curves. Consequently, the shorter-period branch in Figure \ref{fig:PAD} likely consists of low mass-loss AGB stars. Based on the reference region for RSGs, we manually draw the blue line shown in Figure \ref{fig:PAD}, stars to the left of which are excluded from our sample. This process results in a final sample of 6196 late-type Galactic RSG candidates from Gaia RVS spectra, as listed in Table \ref{tab:late_RSGs}.

For these late-type RSG candidates, the $\sigma(\mathrm{EW(CaT)})$ of $\sim$99.4\% sources are lower than 0.07 nm, while $\sim$99.6\% sources have $\sigma(\mathrm{EW(CaT)})/\mathrm{EW(CaT)}$ below 20\%, further demonstrating the reliability of the measurements. In Gaia's {\tt spectraltype\_esphs} classifications, all 6196 stars are labeled as M-type, which is encouraging since the sample is indeed expected to contain only O-rich M-type stars. Again, due to the limitations in Gaia Apsis pipeline of bright late-type stars \citep{2023A&A...671A.148M}, nearly all of the $\sim$1900 stars with available GSP-phot parameters have {\tt logposterior\_gspphot} values below -1000, indicating that their GSP-phot parameters may be unreliable. Nevertheless, the majority of {\tt teff\_gspphot} values fall between 3200-3800K ($\sim$99\%), and {\tt logg\_gspphot} values range from -0.4 to 0.9 ($\sim$99\%), which are consistent with the expected characteristics of RSGs. Among the 72 stars with APOGEE measurements, their $T_{\mathrm{eff}}$ and $\log\ g$ range from 3300-3800K and 0.1-1.2, respectively, further confirming the robustness of the selection criteria.

\section{Disscussion} \label{sec:discussion}

\subsection{The known RSGs} \label{sec:comparison}
Among the 185 known Galactic RSGs mentioned in Section \ref{sec:CaT_of_RSGs}, 14 have RVS spectra with SNR $>$ 100, and four of them are identified as late-type RSGs in our analysis. Of the remaining stars, two lack $W3$-band photometry, while eight have $K-W3$ color $>$ 0.5. In addition, a cross-match with the RSGs identified by \citet{2019AJ....158...20M} yield 23 stars, 15 of which are absent in the 185 known RSGs. Among these 15 stars, one of them lies to the left of the blue boundary in Figure \ref{fig:CaT}, and another has the XP spectra characteristics of S-type stars, both of which \citet{2019AJ....158...20M} classified into the F-region, suggesting they are unlikely to be RSGs. Ten of the remaining stars have $K-W3 > 0.5$, leaving three confirmed as late-type RSG candidates in our sample. The RSGs retained and removed in this work are shown in Figure \ref{fig:candidates} as purple (22) and blue crosses (7), respectively. None of the sources listed in other well-known Galactic RSG catalogs are present in the initial sample of this study, i.e. no RVS spectrum with SNR $>100$ in Gaia DR3.

\subsection{Completeness and pureness} \label{sec:completeness}

\begin{figure*}[h]
	\centering
    \includegraphics[width=0.6\textwidth]{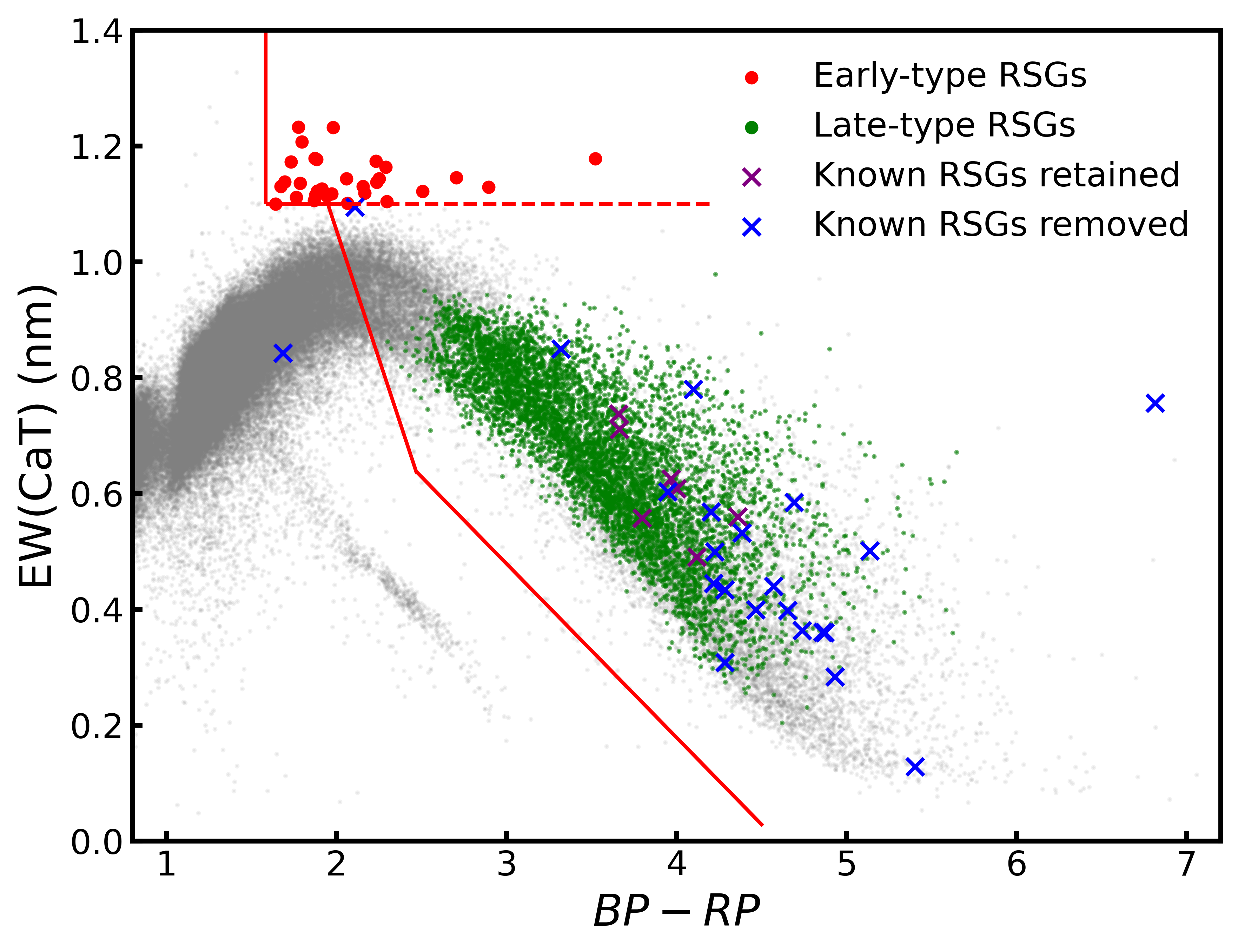}
	\caption{The distribution of RSG candidates selected in this work in the EW(CaT) vs. $BP-RP$ diagram. The red and green dots are the early-type and late-type candidates, respectively. The purple and blue crosses are known RSGs that retained and removed in this work, respectively.  \label{fig:candidates}}
\end{figure*}

This study identified 30 early-type and 6196 late-type RSG candidates, respectively. Their distribution in the EW(CaT) vs. $BP-RP$ diagram is shown in Figure \ref{fig:candidates}. For early-type RSG candidates, the criteria of $BP-RP > 1.584$ and EW(CaT) $>$ 1.1 nm ensure a sufficiently complete sample from the released RVS spectra, as these are fundamental characteristics of RSGs. However, potential contamination may arise from yellow supergiants, which share similar EW(CaT) values and slightly bluer $BP-RP$. If subject to significant extinction, these yellow supergiants could fall within our selection. Another influencing factor is the uncertainty in the EW(CaT) measurements. Taking into account the 6\% relative error described in Section \ref{sec:early_type_RSGs}, under the most stringent conditions, only stars with EW(CaT) $> 1.166$ nm (i.e., 1.1 $\times $ 106\%) can be identified as early-type RSG candidates, resulting in 8 such sources. Under the most lenient conditions, an EW(CaT) $>1.034$ nm would be sufficient to consider a star an RSG candidate, leading to 394 sources.

For late-type RSGs, the requirement of EW(TiO) $>$ 10 nm when analyzing the XP spectra excludes a portion of sources near $BP-RP \sim 2.5$, as evidenced by the absence of candidates in this region in Figure \ref{fig:candidates}. Known RSGs from previous studies do occupy this region, as shown by the blue dots in the left panel of Figure \ref{fig:CaT}, indicating possible incompleteness in our sample. However, as described in Section \ref{sec:TiO}, late-type red giants are present in this area, and EW(TiO) reflects stellar surface gravity, supporting the exclusion of stars with EW(TiO) $<$ 10 nm to avoid contamination from red giants. This decision reflects a balance between pureness and completeness, where pureness takes priority.

In addition, the use of the $K-W3 < 0.5$ criterion likely results in the exclusion of some high-luminosity RSGs, which also possess abundant circumstellar dust caused by significant mass loss. This is indicated by the lack of sources in the lower right corner of Figure \ref{fig:candidates}, where such RSGs are expected to reside. The clustering of blue crosses at the red end in Figure \ref{fig:candidates} confirms this again, as they are likely late-type high-luminosity RSGs. No constraints on photometric quality are used because sources with poorer photometric quality tend to be brighter, making them more likely to be RSGs. This decision is taken to enhance the completeness of our sample, while it may bring about some AGB stars to contamination. By incorporating additional diagnostics from the mid-infrared or far-infrared, it may be possible to improve the sample's pureness further. Again, this also ensures a pure and relatively complete sample of late-type RSG candidates.

\subsection{Application to a larger sample} 

By slightly lowering the SNR threshold to SNR $>$ 50, the total number of available RVS spectra increases to  $\sim$330,000. Applying the process described in Section \ref{sec:RSG} to these spectra results in the identification of 48 early-type and 11,491 late-type RSG candidates. Although the initial sample size has tripled, the final number of RSG candidates has only doubled. This is expected, as the SNR is correlated with apparent magnitude, meaning that lower SNR sources are likely fainter or of lower photometric quality. It is worth noting that the released RVS spectra do not provide uniform coverage across the sky \citep{2023A&A...674A...1G}, with noticeably fewer observations in the Galactic plane. Consequently, many RSGs are missed in our sample. We anticipate that the release of the complete RVS spectra in Gaia DR4 will provide a much more comprehensive perspective for identifying RSGs in the Milky Way.

\section{Summary} \label{sec:summary}

RSGs play a crucial role in processes such as interstellar dust production and massive star evolution, making their identification vitally important. While samples of RSGs in nearby galaxies have been expanding greatly thanks to more precise observational data and more effective methods for excluding foreground sources, the search for RSGs within the Milky Way has been slower. This is partly due to the challenges in applying efficient photometric methods, which are hindered by the difficulties in accurately measuring extinction and distance within our galaxy. Moreover, spectroscopic observations have been too inefficient to yield large samples of RSGs. The release of approximately one million RVS spectra in Gaia DR3 offers new possibilities for the search for Galactic RSGs from spectral features.

In this study, high signal-to-noise ratio (SNR greater than 100) RVS spectra are selected as the initial sample. Firstly, the early-type stars are excluded by using the hydrogen Paschen 14 line. Then, by combining the intrinsic color $BP-RP$ of RSGs across different spectral types with the EW(CaT), the bluest boundary of Galactic RSGs on the EW(CaT) versus $BP-RP$ diagram can be defined. This allows for the selection of early-type RSGs with $BP-RP > 1.584$ and EW(CaT) $>$ 1.1 nm, while removing dwarf stars and the majority of giants, leaving only late-type stars. The analysis of four TiO bands in the XP spectra helps to identify true O-rich M-type stars, thereby avoiding contamination from C-rich and S-type stars. The equivalent widths of the measured TiO bands are averaged and denoted as EW(TiO). Sources with EW(TiO) $<$ 10 nm are considered late-type red giant branch stars, based on their higher $\log\ g$ and bluer colors. The remaining sources are primarily contaminated by AGB stars, where those with mass-loss and those with low mass-loss are identified by $K-W3 > 0.5$ and period-amplitude sequences, respectively, and are excluded. This process yields a final sample of 30 early-type RSG candidates and 6196 late-type RSG candidates.

Applying this method to the RVS spectra with SNR $>$ 50 results in the identification of 48 early-type RSG candidates and 11,491 late-type RSG candidates. However, due to the uncertainties involved in processing low-SNR spectra, these numbers suffer larger uncertainty. This work serves as a preliminary study in anticipation of Gaia DR4, where the release of a larger volume of spectra is expected to have profound implications for the study of Galactic RSGs.

\begin{acknowledgements}
We would like to thank the anonymous referee for the constructive suggestions that definitely improved this work. We thank Dr. R. Dorda for providing the line measurements of stars in the Perseus arm \citep{2018MNRAS.475.2003D}. This work is supported by the National Natural Science Foundation of China (NSFC) through grants Nos. 12133002, 12203025, and 12373048. National Key R\&D Program of China No. 2019YFA0405503, CMS-CSST-2021-A09 and Shandong Provincial Natural Science Foundation through project ZR2022QA064. H.Z. acknowledges the support of the National Natural Science Foundation of China (grant No. 12203099) and the Jiangsu Funding Program for Excellent Postdoctoral Talent. This work has also made use of data from the surveys by Gaia, APOGEE, 2MASS and WISE.
\end{acknowledgements}

\bibliographystyle{aa}
\bibliography{Spectral_RSG}

\clearpage

\clearpage

\begin{table*}[h]
    \caption{The intrinsic color index $(BP-RP)_0$ of Galactic RSGs and their EW(CaT)}              
    \label{tab:BPRP_CaT}      
    \centering                                      
    \begin{tabular}{c c c c}          
    \hline\hline                        
    Spectral type &$T_{\mathrm{eff}}$ (K)& EW(CaT) (nm) & $(BP-RP)_0^{\mathrm{MIST}}$ \\    
    \hline
        K1 & 4100 & $\geq $1.100  & 1.548   \\
        M2 & 3660 & 1.100 & 1.999  \\      
        M3 & 3605 &0.955 & 2.086  \\
        M4 & 3535 &0.810 & 2.221  \\
        M5 & 3450 &0.665 & 2.469  \\
        M6 & --   &0.520 & --  \\
        M7 & --   &0.375 & --   \\
        M8 & --   &0.230 & --  \\
    \hline                                             
    \end{tabular}
\end{table*}

\begin{sidewaystable*}
    \centering
    \caption{The 30 early-type RSG candidates from Gaia RVS spectra}
    \label{tab:early_RSGs}
    \resizebox{\textwidth}{!}{
    \begin{tabular}{lrrrrrrrrr}
         \hline
         Gaia ID & R.A. & Decl. & $G$ & $BP$ & $RP$ & SNR$^a$ & EW(Ca II 8498) & EW(Ca II 8542) & EW(Ca II 8662) \\
         \hline
         & (J2016) & (J2016) & (mag) & (mag)& (mag)& & (nm) & (nm) & (nm)  \\
         \hline
        Gaia DR3 5853442496285943424 & 217.869998 & -62.854653 &        12.078397 &         14.265091 &         10.742440 &             128.383770 &  0.235$\pm$0.010 &  0.540$\pm$0.019 & 0.403$\pm$0.021 \\
        Gaia DR3 5820436432841319296 & 236.608956 & -68.636116 &         8.634509 &          9.516343 &          7.720005 &             313.891540 & 0.229$\pm$0.004 & 0.541$\pm$0.008 & 0.437$\pm$0.008 \\
        Gaia DR3 5820502850214136832 & 238.263848 & -68.346030 &        11.269479 &         12.143257 &         10.358867 &             102.666626 & 0.213$\pm$0.013 & 0.519$\pm$0.023 & 0.403$\pm$0.026 \\
        Gaia DR3 4103709932814452608 & 278.951953 & -14.427099 &         8.090821 &          9.066613 &          7.124403 &             396.023250 & 0.209$\pm$0.003 & 0.497$\pm$0.006 & 0.408$\pm$0.007 \\
        Gaia DR3 5984880628866213760 & 241.238836 & -47.051229 &         8.543653 &          9.319298 &          7.679147 &             295.434970 & 0.213$\pm$0.004 & 0.494$\pm$0.008 & 0.394$\pm$0.009 \\
        Gaia DR3 4093865043071455744 & 272.231815 & -21.402472 &         7.354873 &          8.524316 &          6.293784 &             583.243000 & 0.222$\pm$0.002 &  0.530$\pm$0.004 & 0.422$\pm$0.005 \\
        Gaia DR3 5239689818579893632 & 160.344604 & -65.101892 &         9.030067 &         10.026202 &          8.054391 &             182.615830 &  0.210$\pm$0.007 & 0.508$\pm$0.013 & 0.399$\pm$0.015 \\
        Gaia DR3 4067892929213072512 & 265.895181 & -25.549597 &        10.125677 &         11.799958 &          8.905755 &             169.187670 & 0.214$\pm$0.008 & 0.513$\pm$0.014 & 0.402$\pm$0.016 \\
        Gaia DR3 6073909597165776640 & 193.698323 & -54.162952 &        11.149422 &         12.276551 &         10.111722 &             128.280530 &  0.215$\pm$0.010 & 0.514$\pm$0.019 &  0.390$\pm$0.021 \\
        Gaia DR3 5337904416304099200 & 165.763401 & -61.519718 &        10.441664 &         11.438498 &          9.459789 &             119.273990 & 0.237$\pm$0.011 &  0.551$\pm$0.020 & 0.444$\pm$0.022 \\
        Gaia DR3 5853509983176878592 & 217.053019 & -62.649578 &        10.917154 &         12.289372 &          9.783605 &             201.837100 & 0.216$\pm$0.007 & 0.511$\pm$0.012 & 0.395$\pm$0.013 \\
        Gaia DR3 5853511220127551744 & 217.048239 & -62.579237 &        11.007373 &         12.223262 &          9.928592 &             122.616776 & 0.213$\pm$0.011 &  0.498$\pm$0.020 & 0.393$\pm$0.022 \\
        Gaia DR3 5233812168691960192 & 178.595879 & -70.002198 &         8.382010 &          9.311263 &          7.439888 &             443.421300 & 0.226$\pm$0.003 & 0.526$\pm$0.005 & 0.427$\pm$0.006 \\
        Gaia DR3 5235601284598319616 & 178.660747 & -68.178623 &        10.442018 &         11.313028 &          9.537476 &             104.256640 & 0.229$\pm$0.013 & 0.556$\pm$0.023 & 0.447$\pm$0.025 \\
        Gaia DR3 4102947146590069760 & 279.198009 & -15.945464 &         6.157430 &          7.278494 &          5.123030 &             527.787800 & 0.215$\pm$0.003 &  0.510$\pm$0.005 & 0.405$\pm$0.005 \\
        Gaia DR3 4121402517949300480 & 263.468978 & -20.152666 &         8.579352 &          9.793996 &          7.504721 &             451.467830 & 0.223$\pm$0.003 &  0.520$\pm$0.005 & 0.421$\pm$0.006 \\
        Gaia DR3 4042693875003175808 & 272.360644 & -33.234316 &         6.880441 &          7.680243 &          6.007875 &             489.989320 & 0.211$\pm$0.003 & 0.509$\pm$0.005 & 0.411$\pm$0.005 \\
        Gaia DR3 5636119113908970752 & 137.732247 & -28.473395 &         9.551785 &         10.490046 &          8.603405 &             296.026820 & 0.209$\pm$0.005 & 0.511$\pm$0.008 & 0.403$\pm$0.009 \\
        Gaia DR3 5267828417053112576 & 110.118788 & -69.769439 &        10.266059 &         11.085389 &          9.391768 &             135.958080 &  0.213$\pm$0.010 & 0.522$\pm$0.018 &  0.403$\pm$0.020 \\
        Gaia DR3 867070063597368064 & 115.243793 &  23.018474 &         5.330736 &          6.187736 &          4.425474 &             441.684400 & 0.208$\pm$0.003 & 0.506$\pm$0.005 & 0.397$\pm$0.006 \\
        Gaia DR3 5606087946667386624 & 108.496223 & -30.004131 &        10.214938 &         11.056752 &          9.324802 &             165.470350 & 0.219$\pm$0.008 & 0.529$\pm$0.015 & 0.424$\pm$0.016 \\
        Gaia DR3 4068970240056037504 & 267.321884 & -22.491656 &        10.703414 &         11.622157 &          9.754555 &             122.038020 & 0.215$\pm$0.011 &  0.494$\pm$0.020 & 0.397$\pm$0.022 \\
        Gaia DR3 5845475886035405312 & 198.814510 & -67.919802 &        10.728763 &         11.665064 &          9.783031 &             110.186510 & 0.223$\pm$0.012 & 0.528$\pm$0.022 & 0.426$\pm$0.024 \\
        Gaia DR3 4058702248983652096 & 262.457118 & -29.780555 &         9.034603 &          9.963396 &          8.086934 &             150.555650 & 0.216$\pm$0.009 & 0.499$\pm$0.016 & 0.401$\pm$0.018 \\
        Gaia DR3 5514740550687176960 & 124.620115 & -50.481626 &         9.321692 &         10.272155 &          8.359034 &             235.037890 & 0.216$\pm$0.006 &  0.504$\pm$0.010 & 0.406$\pm$0.011 \\
        Gaia DR3 6004968534471349376 & 226.521595 & -40.550913 &        10.425404 &         11.590588 &          9.354713 &             105.546530 & 0.188$\pm$0.013 & 0.522$\pm$0.023 & 0.427$\pm$0.025 \\
        Gaia DR3 6723841434427244416 & 278.550854 & -39.303318 &         9.025936 &         10.073758 &          8.016438 &             273.550870 & 0.214$\pm$0.005 & 0.517$\pm$0.009 &  0.413$\pm$0.010 \\
        Gaia DR3 5515454717849642624 & 127.087196 & -48.134389 &        10.079885 &         11.600745 &          8.895877 &             171.946580 & 0.218$\pm$0.008 & 0.519$\pm$0.014 & 0.408$\pm$0.016 \\
        Gaia DR3 3155188292539070208 & 109.034100 &   8.823780 &         9.506177 &         10.560427 &          8.496337 &             131.214870 &  0.206$\pm$0.010 &   0.500$\pm$0.019 &  0.395$\pm$0.020 \\
        Gaia DR3 5546711192039738752 & 123.739687 & -33.449050 &         8.767647 &          9.926933 &          7.676443 &             384.052580 & 0.217$\pm$0.003 & 0.509$\pm$0.006 & 0.418$\pm$0.007 \\
         \hline
    \end{tabular}
    }
\end{sidewaystable*}

\begin{sidewaystable*}
    \centering
    \caption{The 6196 late-type RSG candidates from Gaia RVS spectra}
    \label{tab:late_RSGs}
    \resizebox{\textwidth}{!}{
    \begin{tabular}{lrrrrrrrrrrrr}
         \hline
         Gaia ID & R.A. & Decl. & $G$ & $BP$ & $RP$ & SNR & EW(Ca II 8498) & EW(Ca II 8542) & EW(Ca II 8662) & $K$ & $W3$ & EW(TiO) \\
         \hline
         & (J2016) & (J2016) & (mag) & (mag)& (mag)& & (nm) & (nm) & (nm) & (mag) & (mag) & (nm) \\
         \hline
         Gaia DR3 5912905292012078080 & 263.831986 & -58.796211 &        10.944374 &         13.220857 &          9.563824 &             164.047710 &  0.062$\pm$0.009 & 0.252$\pm$0.016 & 0.209$\pm$0.018 & 5.899 &  5.480 &       17.653 \\
        Gaia DR3 5912586227479692800 & 262.643347 & -59.522606 &        10.615781 &         13.111814 &          9.200260 &             155.498920 &   0.045$\pm$0.010 & 0.234$\pm$0.017 & 0.169$\pm$0.019 & 5.439 &  5.209 &       20.464 \\
        Gaia DR3 2269271710582159232 & 277.967082 &  75.698835 &         9.047530 &         10.693685 &          7.799423 &             244.999390 &  0.152$\pm$0.006 &  0.392$\pm$0.010 & 0.327$\pm$0.011 & 4.730 &  4.580 &       12.248 \\
        Gaia DR3 2265488703388096896 & 283.316176 &  70.711737 &        10.098157 &         11.943086 &          8.815959 &             217.533920 &  0.124$\pm$0.007 & 0.343$\pm$0.012 & 0.294$\pm$0.013 & 5.608 &  5.431 &       15.031 \\
        Gaia DR3 5902334724683114880 & 227.324252 & -49.648354 &        11.454111 &         14.084639 &         10.028410 &             101.382140 &  0.071$\pm$0.015 & 0.262$\pm$0.027 & 0.202$\pm$0.029 & 5.998 &  5.782 &       19.736 \\
        Gaia DR3 5902417703445930624 & 228.320978 & -49.174342 &        10.891149 &         13.215270 &          9.530221 &             120.433140 &  0.105$\pm$0.012 & 0.306$\pm$0.022 &  0.260$\pm$0.024 & 5.820 &  5.557 &       16.115 \\
         \hline
    \end{tabular}
    }
    \tablefoot{(This table is available in its entirety in machine-readable form.)}
\end{sidewaystable*}

\clearpage

\end{CJK*}
\end{document}